\documentclass[%
 reprint,
superscriptaddress,
 amsmath,amssymb,
 aps,
prb,
floatfix,
fleqn
]{revtex4-2}

\usepackage{xcolor}
\usepackage{graphicx}
\usepackage{dcolumn}
\usepackage{bm}
\usepackage[colorlinks=true,citecolor=blue,linkcolor=magenta]{hyperref}
\usepackage{braket}

\usepackage{cases}
\usepackage{soul}
\usepackage[normalem]{ulem}

\def \mc{\mathcal}

\begin{document}


\title{Spin-dependent anisotropic electron-phonon coupling in KTaO$_3$}
\author{Giulia Venditti}%
\affiliation{Department of Quantum Matter Physics, University of Geneva, 24 Quai Ernest-Ansermet, 1211 Geneva, Switzerland}%
\author{Francesco Macheda}
\affiliation{Dipartimento di Scienze e Metodi dell’Ingegneria, University of Modena and Reggio Emilia, Reggio Emilia, Italy}
\author{Paolo Barone}
\affiliation{SPIN-CNR Institute for Superconducting and other Innovative Materials and Devices, Area della Ricerca di Tor Vergata, Via del Fosso del Cavaliere 100, 00133 Rome, Italy}
\author{Jos\'e Lorenzana}
\affiliation{ISC-CNR Institute for Complex Systems and Department of Physics, Sapienza University of Rome, Piazzale Aldo Moro 2, 00185, Rome, Italy}
\author{Maria N. Gastiasoro}
\email{maria.ngastiasoro@dipc.org}
\affiliation{Donostia International Physics Center, 20018 Donostia-San Sebastian, Spain}

\date{\today}
             
\begin{abstract}
KTaO$_3$ (KTO) is an incipient ferroelectric, characterized by a softening of the lowest transverse optical (TO) mode with decreasing temperature. 
Cooper pairing in the recently discovered KTO-based heterostructures has been proposed to be mediated by the soft TO mode.
Here we study the electron coupling to the zone-center odd-parity modes of bulk KTO by means of relativistic Density Functional Perturbation Theory (DFPT). The coupling to the soft TO mode is by far the largest, with comparable contributions from both intraband and interband processes. Remarkably, we find that for this mode, non-spin-conserving matrix elements are particularly relevant. We develop a three-band microscopic model with spin-orbit coupled $t_{2g}$ orbitals that reproduces the main features of the \emph{ab initio} results. For the highest energy band, the coupling can be understood as a ``dynamical" isotropic Rashba effect. In contrast, for the two lowest bands, the Rashba-like coupling becomes strongly anisotropic.  
The DFPT protocol implemented here enables the calculation of the full electron-phonon coupling matrix projected onto any mode of interest, and it is easily applicable to other systems.
\end{abstract}

\maketitle

\section{Introduction}
\label{sec:intro}


The anomalous lattice dynamics and dielectric properties of SrTiO$_3$ (STO) and  KTaO$_3$ (KTO) have puzzled the condensed matter community for many decades. Their inverse dielectric susceptibility decreases linearly with temperature, extrapolating to a putative Curie temperature at which a transition to a ferroelectric (FE) state should occur. 
Instead, at low temperatures, the susceptibility saturates and the system remains paraelectric \cite{Barret1952,wemple1965}.
Concomitantly, both STO and KTO show the softening of a transverse optical (TO) mode \cite{Cowley1964,Axe1970,Migoni1976} without condensation into the inversion symmetry-breaking FE state. This has been explained in terms of quantum fluctuations leading to the concept of a quantum paraelectric \cite{Muller1979,rowley2014}, and very recently, in terms of lamellar fluctuations arising from the coupling between the soft TO mode and strain gradients \cite{guzman2023}. 

The introduction of mobile electrons leads to a dilute metallic state with sharp Fermi surfaces (FS) \cite{lin2013} and interesting physics.  
Indeed, doped STO was the first material to show a superconducting dome upon doping~\cite{Koonce1967}, and despite many decades of research, the superconducting state still defies our understanding \cite{gastiasoro2020review}. The discovery of gate-tunable superconductivity in STO interfaces and heterostructures with other insulators opened a route to study its two-dimensional limit~\cite{Caviglia2008,pai2018}. The presence of a dome and similar transition temperatures $T_c\sim 0.2$\,K suggests they may foster the same pairing mechanism. More recently, a new platform for superconductivity was found on the related interfaces between KTO and other oxide insulators, as well as on uncapped KTO surfaces doped with ionic gating \cite{ueno2011,liu2021two,cheng2021,chen2021electric,ren2022two,liu2022tunable,mallik2022superfluid,maryenko2023,chen2024orientation}. The one order of magnitude higher $T_c$ in some of the latest interfaces, up to $\sim 2$ K, is not only remarkable but has also the potential to shed light on the pairing mechanism in doped quantum paraelectrics. 

The existence of the soft TO mode led to pairing proposals invoking FE fluctuations \cite{edge2015,Wolfle2018a,kozii2019,gastiasoro2020}, and extensive studies on STO have reported sensitivity of $T_c$ to the putative FE quantum critical point (QCP) \cite{rischau2017,Rischau2022,tomioka2022,hameed2022,saha2025}.   
A crucial ingredient in this electron-boson model is the coupling of the itinerant electrons to the soft mode, since transverse modes do not produce long-range electric fields; therefore, the conventional Fr\"oelich electron-phonon coupling vanishes in the infrared limit \cite{ruhman2019}.
Interesting models to overcome this issue have been proposed \cite{gastiasoro2020review}, including a two-phonon coupling \cite{Ngai1974,VanderMarel2019,Kiselov2021}  and a linear Rashba-like EPC allowed in the presence of spin-orbit coupling (SOC) \cite{kozii2015odd, gastiasoro2022theory,yu2022,gastiasoro2023,klein2023}. 
The latter has also been discussed in connection with collective modes in the so-called polar metal phase  \cite{volkov2023} and similar ideas have been invoked in other contexts, such as establishing spin textures by driving coherent phonons \cite{Schlipf2021}. 
At present, there is no consensus on the mechanism of superconductivity in STO, for either bulk or heterostructures. 

The recently discovered interface superconductivity in KTO has the peculiarity of a strong crystallographic orientation dependence of $T_c$, with a maximum for [111] oriented interfaces up to $T_c\approx 2\,$K (an order of magnitude higher than in STO), $T_c\approx 1\,$K in [110], and no sign of superconductivity in the [001] down to 25 mK.   

Because of the heavy Ta ion, KTO also presents a strong SOC, splitting the $t_{2g}$ orbitals at $\Gamma$ by $\approx 400\,$meV \cite{Mattheiss1972,uwe1980raman}, a 14 times larger split than in STO. Moreover, a strong violation of the Pauli limit for in-plane magnetic fields is observed both in [111] and [110] oriented interfaces \cite{arnault2023,al-tawhid2023,filippozzi2024,poage2025violation,yang2025}, an effect that has been attributed to SOC effects. 
Interestingly, the softening of the TO mode has also been recently measured in STO- and KTO-based interfaces by surface-sensitive spectroscopy  \cite{chu2024}. All these facts raise the question of the relevance of the SOC-assisted Rashba-like coupling to the soft mode for superconductivity, and may offer new insights into the pairing mechanism. 

In the insulating regime, the zero-$q$ longitudinal optical  (LO) component stiffens due to the long-range Coulomb interaction, making the soft mode transverse. At low densities, when the electronic plasma frequency is lower than the LO phonon frequency, the LO-TO splitting remains. At sufficiently high densities, however, the electronic screening causes both frequencies to become degenerate, eventually rendering them both soft near the FE instability. In this case, the momentum dependence becomes important, as the screening occurs only for wavevectors smaller than the Thomas-Fermi wavevector. Taking into account the screening of both the matrix elements and the phonon frequency is challenging and requires a very high momentum resolution~\cite{PhysRevLett.129.185902}.

Because in the conventional EPC (i.e., without spin-orbit effects), the coupling to the LO phonon dominates, how the LO frequency is treated is crucial. Ref.~\cite{esswein2022first} considered both conventional EPC and spin-orbit effects, but, as the authors acknowledge, did not take fully into account the LO-TO splitting. This may lead to overestimating the conventional EPC as the frequency of the LO phonon is assumed to be soft in a large region of momentum space. Hence,  it becomes difficult to compare the conventional and SOC-assisted EPC on the same grounds.     

Here, we focus on the coupling to the soft mode through the ``dynamical" Rashba effect, which is expected to display a weak dependence on momentum and doping, in contrast to the conventional coupling to the LO mode, which requires very demanding numerical resources (i.e. doping-dependent large momentum resolution), beyond our present scope.
	
The Rashba-like coupling originates from real-space electronic hopping processes induced by the atomic displacements of the phonons. In a simplified model for KTO, they can be understood in terms of hopping processes among spin-1/2 $t_{2g}$ orbitals $\ket{\mu\sigma}$ with $\mu=x,y,z$, corresponding to $d_{zy},d_{zx}$ and $d_{zx}$ orbitals respectively.
For {\em spin-conserving} hopping processes, the EPC matrix element results of order  $t_{\mu\nu}$. Here $t_{\mu\nu}$ is the derivative of a hopping matrix element between two orbitals in neighboring sites with respect to a polar distortion, times the harmonic characteristic length of the phonon, and thus has units of energy. 
Moreover, SOC itself combined with a lattice distortion can induce {\em spin-non-conserving} hopping processes among $t_{2g}$ orbitals directly or indirectly due to mixing with other orbitals (e.g. via the $e_g$ orbitals of Ta or the $p$ orbitals of O) and orbital polarization effects (e.g. $d$-orbital deforms acquiring $f$-orbital component).   
We will refer to these distinct processes as Rashba type I (spin-conserving) and Rashba type II (spin-non-conserving).
We already speculated in Ref.~\cite{venditti2023} that Rashba type II processes may play an important role in the coupling to the soft mode in KTO. In this work, we go beyond the methods in Ref.~\cite{venditti2023}, analyzing type I and type II processes on an equal footing.

We study the EPC to the zone-center modes of KTO by comparing relativistic Density Functional Perturbation Theory (DFPT) with a three-band model of spin-orbit coupled $t_{2g}$ orbitals.
Going beyond {\em frozen phonon} computations, we compute the total electron-phonon matrix resolved in pseudospin and 
projected onto KTO zone center eigenmodes. All modes show comparable contributions from intraband and interband processes, and the coupling to the soft TO mode is found to be the largest. For this mode, Rashba type II processes are particularly relevant. The coupling to this mode induces dynamical Rashba-like isotropic chiral angular momentum currents to the highest electronic band. In contrast, for the two lowest bands, the coupling becomes strongly anisotropic, beyond the conventional Rashba effect. 
The DFPT protocol presented here allows for the extraction of coupling coefficients for a given vibrational eigenvector, including spin effects, and it is easily applicable to other systems.

The paper is organized as follows: in Section~\ref{sec:generalEPC}, we present the relevant electronic bands and zone-center phonon modes in KTO and introduce the EPC using symmetry arguments;
our DFTP protocol and calculations are presented in Section~\ref{sec:DFPT}; the three-band microscopic model that reproduces most of the DFPT results is then discussed in Section~\ref{sec:model}, where we demonstrate the importance of non-spin-conserving processes in the EPC to the soft mode;
we finally draw our conclusions in Section~\ref{sec:conclusions}.

\section{Electrons, Phonons and their coupling}
\label{sec:generalEPC}

\begin{figure}
    \centering
    \includegraphics[width=\columnwidth]{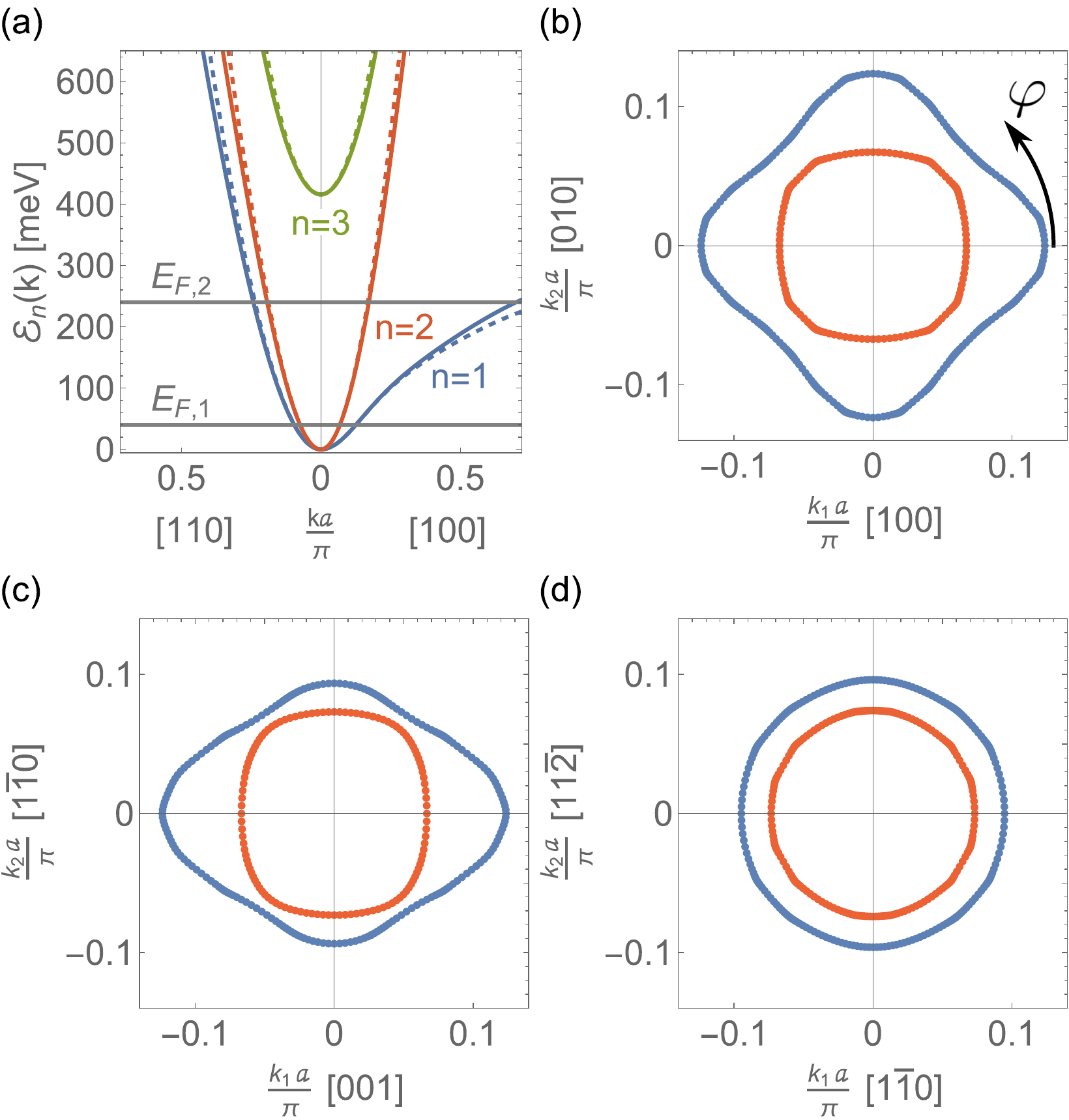}
    \caption{
    \emph{Electronic band structure of KTaO$_3$.}
    (a) Full lines are the \emph{ab initio} results of the three conduction bands along M$-\Gamma-$X. Dashed lines are the tight-binding model [Eq~\eqref{eq:H0}] with the $3\xi=416$ meV SOC gap at the zone center. Gray lines are at Fermi energies $E_{F,1}=40$ meV and $E_{F,2}=240$ meV. The Fermi surface for $E_{F,1}$ with $n=1$ outer band (blue) and $n=2$ inner band (red), in the plane $(k_1,k_2,0)$ perpendicular to (b) [001], (c) [110] and (d) [111]. The angle in panel (b) is used in Fig.~\ref{fig:epc-phi}, Fig.~\ref{fig:qe-vs-tb-tu} and Fig.~\ref{fig:qe-vs-tb}.
    }
    \label{fig:FS}
\end{figure}

\subsection{Electronic properties}
The electronic structure of KTO near the Fermi level comes mainly from the 5d $t_{2g}$ electrons of Ta \cite{bruno2019band}, with a strong SOC splitting of the bands of around $400\,$meV.
In terms of the spinors of the three spin-$1/2$ $t_{2g}$ orbitals  $\varphi^\dagger_\mu(\bm k)=(c^\dagger_{\mu+}(\bm k),c^\dagger_{\mu-}(\bm k))$ (with orbital index $\mu=x,y,z$) the low-energy part of the relativistic \emph{ab initio} electronic dispersion [full lines in Fig.~\ref{fig:FS}(a)] can be described~\cite{venditti2023} by the following tight-binding Hamiltonian,
\begin{flalign}
\label{eq:H0}
\mathcal{H}_0&=\sum_{\bm{k}}\sum_{\mu\nu} \varphi^\dagger_\mu(\bm k) t^{(0)}_{\mu\nu}(\bm k)\sigma_0\varphi_\nu(\bm k)+i\xi\left[ \varphi^\dagger_x(\bm k)\sigma_3\varphi_y(\bm k)\right.&\\\nonumber
&\left.+\varphi^\dagger_y(\bm k)\sigma_1\varphi_z(\bm k)+\varphi^\dagger_z(\bm k)\sigma_2\varphi_x(\bm k)\right]+\mathrm{h.c.}&\\
&=\sum_{\bm{k}}\sum_n\psi_n^\dagger(\bm k) \mathcal{E}_n(\bm k)\sigma_0\psi_n(\bm k)&\label{eq:H0soc}
\end{flalign}
with the 2$\times$2 identity matrix $\sigma_0$, ${\bm \sigma}\equiv(\sigma_1,\sigma_2,\sigma_3)$, and Pauli matrices $\sigma_i$ acting on the spinors. 
Here $t^{(0)}_{\mu\nu}(\bm k)$ are hopping terms between $t_{2g}$ orbitals $\mu$ and $\nu$ up to next-nearest neighboring Ta atoms, and $\xi$ is the effective atomic SOC (see Appendix~\ref{app:appH0} for details). These parameters have been extracted by a fit to the relativistic \emph{ab initio} electronic structure and result in three two-fold degenerate electronic bands $\mathcal{E}_n(\bm k)$ (dashed lines in Fig.~\ref{fig:FS}(a)) in terms of the SOC electronic spinors $\psi_n^\dagger(\bm k)=(c^\dagger_{n+}(\bm k),c^\dagger_{n-}(\bm k))$ with pseudospin $\pm$. We will henceforth refer to the doubly degenerate bands as $n=1,\,2,\,3$ (blue, red, green). 

The transformation from the orbital basis to the band basis is particularly simple close to $\Gamma$.
Because of the SOC term, at the $\Gamma$ point states separate into a $j=3/2$ quartet (bands $n=1,2$) and a $j=1/2$ doublet (band $n=3$). Here $j$ is an effective total angular momentum resulting from the mapping of the three $t_{2g}$ orbitals into $p$ orbitals. Then the effective $l=1$ orbital angular momentum is added to the spin-$1/2$ angular momentum \cite{Stamokostas2018,Khomskii2021,gastiasoro2022theory,gastiasoro2023}. 

The following 
states diagonalize the Hamiltonian $\mathcal{H}_0$  up to linear order in momentum around $\Gamma$, labeled by the effective total angular momentum and its $z$-projection $\{j,j_z\}$,
\begin{flalign}
 &\Bigl\{\frac32,\pm \frac12\Bigr\}:&\nonumber \\ 
&\ \ \ \ \ \  \psi_1^\dagger(\bm k)=\frac1{\sqrt{6}}\left[-i\sigma_2\varphi_x^\dagger(\bm k)-i\sigma_1\varphi_y^\dagger(\bm k)+2\sigma_0\varphi_z^\dagger(\bm k)\right]&
\nonumber
 \\
&\Bigl\{\frac32,\mp \frac32\Bigr\} : &\nonumber\\
& \ \ \ \ \ \ \psi_2^\dagger(\bm k)=\frac1{\sqrt{2}}\left[i\sigma_2\varphi_x^\dagger(\bm k)-i\sigma_1\varphi_y^\dagger(\bm k)\right)]&
\label{eq:fi2psi}
 \\
&\Bigl\{\frac12,\pm \frac12\Bigr\} :&\nonumber\\ 
&\ \ \ \ \ \ \psi_3^\dagger(\bm k)=\frac1{\sqrt{3}}\left[-i\sigma_2\varphi_x^\dagger(\bm k)-i\sigma_1\varphi_y^\dagger(\bm k)-\sigma_0\varphi_z^\dagger(\bm k)\right].&
\nonumber
 \end{flalign}
Here, the pseudospin index was chosen following the same convention as in Ref.~\cite{gastiasoro2023}.  
Note that to linear order in  $\bm k$, there is no $\bm k$ dependence of the coefficients of the transformation. As we shall see, the linear order is enough to obtain the behavior of the EPC near $\Gamma$.

\subsection{Vibrational properties and electron phonon coupling}
The 
electron-phonon Hamiltonian to first order in the atomic displacements for the electronic spinor $\psi_n^\dagger(\bm{k})$
for phonon mode $\lambda$ is:
\begin{flalign}
\label{eq:Hepc}
     &\mc{H}_\text{EPC}=\frac{1}{\mathcal{N}}\sum_{\bm k, nm,\bm q \lambda} \psi_n^\dagger(\bm{k}+\frac{\bm{q}}{2})  \Lambda_{nm,\lambda}(\bm k,\bm q)\psi_m(\bm{k}-\frac{\bm{q}}{2}) \mc{A}_{\bm q,\lambda}&
\end{flalign}
Here $\mc{A}_{\bm q,\lambda}=a_{\bm q,\lambda}+a^\dagger_{-\bm q,\lambda}$ is the phonon operator, and the general EPC matrix reads
\begin{equation}
    \label{eq:Lambda}
    \Lambda_{nm,\lambda}(\bm k,\bm q)= g_{nm,\lambda} (\bm k, \bm{q}) \sigma_0 + \bm{G}_{nm,\lambda} (\bm{k},\bm q)\cdot \bm{\sigma} 
\end{equation}

The pseudospin-independent coupling $g_{nm,\lambda}(\bm k, \bm q)$ is the conventional scalar electron-phonon matrix element. 
The vector 
$\bm{G}_{nm,\lambda}(\bm k, \bm q)$ represents instead a pseudospin-dependent coupling, which has been considered less often in literature. 

In this work, we will carefully study the intraband ($n=m$) and interband ($n\neq m$) EPC $\Lambda_{nm,\lambda} (\bm{k},\bm q)$ by a particular post-processing of electron-phonon coefficients computed with DFPT using QE in KTO. With respect to our previous work \cite{venditti2023}, the advantage of DFPT over frozen phonon is two-fold: 1) \emph{interband} coupling terms can be accessed, and 2) the computation to any $\bm q\neq 0$ can be straightforwardly extended (without the need of large supercells), which we will also explore.

In cubic KTO \emph{all} 15 phonon modes at the zone center are odd-parity modes, with irreducible representations (irreps) $4 T_{1u}\oplus T_{2u}$. 
The three acoustic and nine optical (di)polar modes belong to the $T_{1u}$ irrep ($l=1$), and the three optical octopolar modes to the $T_{2u}$ irrep ($l=3$). The threefold degeneracy of the irreps corresponds to the three possible polarizations. At infinitesimal $\bm q$, the dipolar triplets get further split by long-range Coulomb into one longitudinal mode and two transverse modes. 

In the presence of inversion and time-reversal symmetries, for all these odd-parity zone center modes the pseudospin-independent intraband term in Eq.~\eqref{eq:Lambda} must vanish, $g_{nn,\lambda}(\bm k,\bm q=0)\sigma_0=0$, and the pseudospin-dependent term can be finite instead, $\bm{G}_{nn,\lambda} (\bm{k},\bm q= 0)\cdot \bm \sigma \neq 0$ (for the acoustic modes the coupling is linear in $q$, so it does trivially vanish at $q=0$). It would be desirable to obtain the symmetry allowed form of the non-vanishing intraband coupling $\bm{G}_{nn,\lambda} (\bm{k},\bm q= 0)$ to lowest order in $\bm k$.

The matrix  $\bm{G}_{nn,\lambda} (\bm{k},\bm q= 0)$ generally depends on the chosen basis $\psi_n$. Indeed, 
because of Kramers degeneracy, any $U(2)$ rotation of the basis produces another allowed basis \cite{Blount1985,Fu2015}. Near $\Gamma$, the basis dependence of the matrix elements can be circumvented by expressing the EPC in terms of matrices representing the effective angular momentum operators $\hat j_x,\hat j_y,\hat j_z$.

In the case of $j=1/2$ [$n=3$, Eq.\eqref{eq:fi2psi}] this is particularly simple as the system maps into a spin-1/2 system with angular momentum $\bm j$.
In the $O_h$ point group one can build the following odd-parity objects combining $\bm k$ and $\bm j$: $ T_{1u} \otimes T_{1g}=A_{2u} \oplus E_{u} \oplus T_{1u}\oplus T_{2u}$. 
From the $T_{1u}$ term, we obtain a dynamic Rashba-like coupling to polar modes with polarization $\hat{\bm n}$~\cite{kozii2015odd,kanasugi2018,volkov2020,gastiasoro2020,sumita2020,gastiasoro2022theory},
\begin{equation}
    \bm{G}_{33,T_{1u}^\lambda} (\bm{k},\bm q=\bm 0)\cdot \bm{\sigma}=\bar g_{3,T_{1u}^\lambda} \left(\hat k_x j_y-\hat k_y j_x\right) \hat{n}_z
    \label{eq:G-sym-T1u}
\end{equation}
where the angular momentum matrices $j_i$ are proportional to Pauli matrices. In the presence of a lattice distortion, this form implies a splitting in pseudospin space which is independent of the azimuthal angle $\varphi$ defined in Fig.~\ref{fig:FS}(b). Here, we have defined the $z$ axis to be parallel to the polarization of the phonon $\hat{\bm n}$, without loss of generality. This coupling is maximum (zero) when the electronic momentum $\bm k$ is perpendicular (parallel) to the polar axis of the mode $\hat{\bm n}$.
Analogously, for the $T_{2u}$ term we obtain a dynamic linear Dresselhaus-like coupling 
\begin{equation}
    \bm{G}_{33,T_{2u}} (\bm{k},\bm q=\bm 0)\cdot \bm{\sigma}=\bar g_{3,T_{2u}} \left(\hat{k}_x  j_y+\hat{k}_y j_x\right)\hat{n}_z
      \label{eq:G-sym-T2u}
\end{equation} 
to octopolar modes.

For the $j=3/2$ manifold, within linear in $k$ order, the degeneracy problem is more severe as the states are fourfold degenerate at $\Gamma$ and split only quadratically in $k$ by the kinetic energy term in $\mathcal{H}_0$. In contrast, the phonon perturbation produces a linear-in-$k$ splitting. Therefore, in this limit, the phonon-induced mixing between bands $\psi_1^\dagger$ and $\psi_2^\dagger$ cannot be neglected, and one should consider 4x4 matrices. One can show that the $n=1$ intraband term involves $ j^+= j_x+i j_y$ to the third power and its hermitian conjugate. In general, one can express the perturbation in terms of linear and third powers of $\bm j$ angular momentum matrices. As we shall see, for $j=3/2$, the general form departs from the conventional isotropic Rashba/Dresselhaus forms in that the azimuthal angular dependence of the phonon-induced band splitting becomes important. We call this anisotropic Rashba/Dresselhaus-like EPC~\cite{bruno2019band,venditti2023}.

In the following, we show how to compute the symmetry-allowed EPC matrix elements for the $T_{1u}$ and $T_{2u}$ modes via relativistic DFPT computations in KTO. 

\section{Density Functional Perturbation Theory}
\label{sec:DFPT}

\subsection{Definitions}
\label{sec:dfptdef}
We consider the standard Khon-Sham single-particle Hamiltonian,
\begin{align}
H=T+V_{\rm KS}
\end{align}
where $T$ is the non-interacting kinetic energy and $V_{\rm KS}$ is the Khon-Sham potential containing the Hartree interaction and exchange-correlation effects \cite{Kohn1999}. 
The Hamiltonian is diagonalized by the Bloch functions
\begin{align}
\Psi_{n\eta{\bm k}}({\bm r})=e^{i{\bm k}\cdot {\bm r}}u_{n\eta{\bm k}}({\bm r}),
\end{align}
where $u_{n\eta{\bm k}}({\bm r})$ have the same periodicity of the lattice and $\eta=\pm$ is the spinor index. We then consider a phonon polarization vector $e^{\lambda}_{\kappa\alpha}({\bm  q})$, normalized in the unit cell $p$, for the atom $\kappa$ along the Cartesian direction $\alpha$, which induces the following real-space displacement ${\bm \upsilon}$:
\begin{align}
\upsilon^{\lambda}_{\kappa p,\alpha}({\bm  q}) = e^{\lambda}_{\kappa\alpha}({\bm  q}) e^{i{\bm  q}\cdot({\bm  R}_p+\boldsymbol{ \tau}_\kappa)},
\label{eq:phaseconv}
\end{align}
where ${\bm  R}_p$ is the vector that identifies the cell $p$ and $\lbrace \boldsymbol{ \tau}_\kappa \rbrace$ are the basis vectors of the atoms $\kappa$ in the unit cell. Usually, one defines as the electron-phonon coefficients 
\begin{flalign}
&\Lambda_{nm,\lambda}(\bm k,\bm q)\Big|_{\{\eta,\eta'\}}=\langle u_{n\eta{\bm k+\bm q/2}}|\Delta_{{\bm q}\lambda} v_{\rm{KS}}| u_{m\eta'{\bm k-\bm q/2}} \rangle&
\end{flalign}
where $\Delta_{{\bm q},\lambda} v_{\rm KS}$ is the cell-periodic variation of the Khon-Sham potential, induced by the displacement of the atom  $\kappa$ along the Cartesian direction $\alpha$ as
\begin{align}
\Delta_{{\bm  q}\lambda} v_{\rm KS}&=e^{-i{\bm  q}\cdot {\bm r}}\Delta_{{\bm  q}\lambda} V_{\rm KS},\\
\Delta_{{\bm  q}\lambda} v_{\rm KS}&=\sum_{\kappa\alpha}\sqrt{\frac{\hbar}{2M^{\kappa}\omega_{{\bm  q}\lambda}}}e^{\lambda}_{\kappa\alpha}({\bm  q})\partial_{{\bm  q}\kappa\alpha}v_{\rm KS},\\
\partial_{{\bm  q}\kappa\alpha}v_{\rm KS}&= \sum_{p}e^{-i{\bm  q}\cdot ({\bm  r}-{\bm  R}_p-\boldsymbol{\tau}_\kappa)}\frac{\partial V_{\rm KS}}{\partial  (R_{p\alpha}+\tau_{\kappa\alpha})}
\end{align}
where $M^{\kappa}$ is the mass of the atom with $\kappa=\mathrm{Ta,K,O_x,O_y,O_z}$. Here, O$_\mu$ indicates the three oxygens in the unit cell with identical mass $M^\mathrm{O}$.

For the purpose of this work, it is more convenient to compute first the coefficients
\begin{flalign}
\label{eq:gamma}
 &\gamma_{nm}^{\kappa\alpha}(\bm k,\bm q)\Big|_{\{\eta,\eta'\}}=\langle u_{n\eta{\bm k+\bm q/2}}|\partial_{{\bm q}\kappa\alpha} v_{\rm{KS}}| u_{m\eta'{\bm k-\bm q/2}} \rangle &
\end{flalign}
which, for a given set of bands $n$ and $m$, is a 2$\times$2 matrix. Then, we recover the electron-phonon coupling as a 6$\times$6 matrix via
\begin{flalign}
&\Lambda_{nm,\lambda}(\bm k,\bm q)=\sum_{\kappa\alpha}\sqrt{\frac{\hbar} {2 \omega_{\bm q\lambda} M^{\kappa}}}\gamma^{\kappa\alpha}_{nm} (\bm k,\bm q) e^{\lambda}_{\kappa\alpha}(\bm q),&
\label{eq:LambdaQE0}
\end{flalign}
in terms of phonon eigenvectors $e^{\lambda}_{\kappa\alpha}(\bm q)$.
In this way, we divide the task of finding the electron phonon interaction into two sub-tasks that can be performed independently: i) compute the matrix elements Eq.~\eqref{eq:gamma} via DFPT ii) find, from first principles or experimentally, the phonon eigenvectors and frequencies to insert in Eq.~\eqref{eq:LambdaQE0}.

This procedure guarantees a finer control of the electron-phonon coupling, as one can decide afterwards which is the more appropriate phonon eigenvector for a given mode. This is useful in incipient ferroelectrics, as it may not be a priori obvious what phonon eigenvector to use due to intrinsic anharmonicities. We present a simple choice below, but our procedure can be generalized to other choices of phonon eigenvectors.

 \textit{Degeneracies---} 
The electron-phonon coupling of Eq. \eqref{eq:LambdaQE0} is expected to contain pseudospin-dependent terms in the form of Eqs.~\eqref{eq:G-sym-T1u}-\eqref{eq:G-sym-T2u}. The \textit{ab initio} $\gamma^{\kappa\alpha}_{nm}$ in Eq.~\eqref{eq:gamma} are determined using the ground state unperturbed Bloch wavefunctions, which present degenerate subspaces $\mathcal{D}_{m\bm{k}}$ and $\mathcal{D}_{n\bm{k+q}}$ 
($\mathcal{D}_{m\bm{k}}\ge2$ due to Kramer's theorem and inversion symmetry). 
This introduces a gauge freedom in the choice of the wavefunctions; therefore, matrix elements in general depend on this choice of gauge.
For intraband $\bm q=0$ matrix elements, we can characterize the electron phonon coupling by reporting the splitting of the band for a given displacement, which is clearly gauge invariant. For all other cases (finite $\bm q$ or  interband matrix elements), we will use the gauge-invariant quantity, 
\begin{equation}
\mathfrak{g}_{nm,\lambda}(\bm k,\bm q)=\sqrt{\sum_{\substack{m' \in \mathcal{D}_{m{\bm k}} \\ n' \in \mathcal{D}_{n{\bm k+\bm q}}}}\frac{|\Lambda_{n'm',\lambda}(\bm k,\bm q)|^2}{2}}
\label{eq:avrg}
\end{equation} 
for the same purpose.

\subsection{Zone center modes (q=0)}

\begin{figure*}
    \centering
\includegraphics[width=\textwidth]{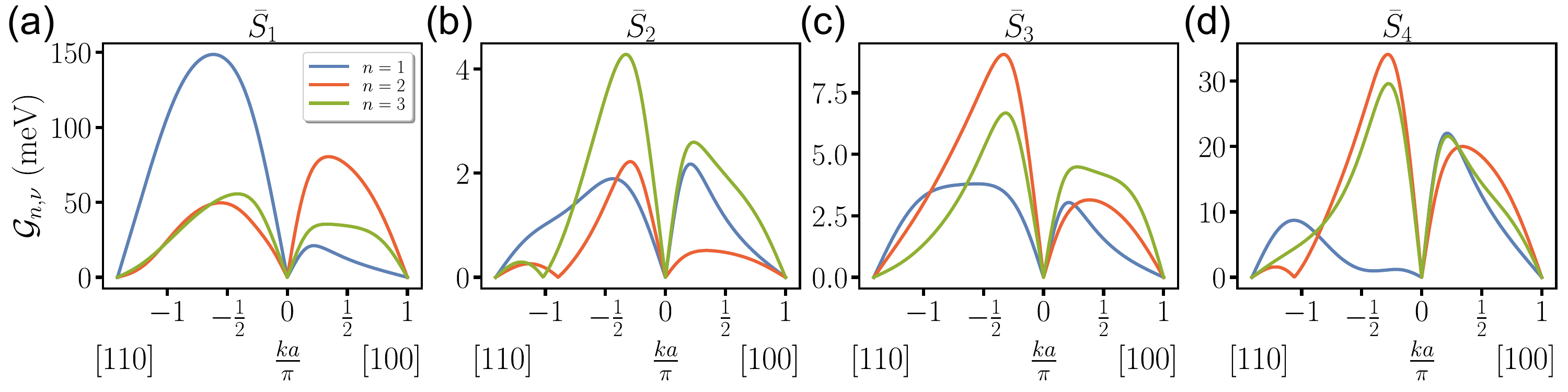}
    \caption{
    \emph{Intraband EPC to zone-center odd-parity modes polarized along $[001]$ computed by QE.} 
    $\mc{G}_{n,\lambda}(\bm k)$ in Eq.~\eqref{eq:QE-intra} for the electronic band $n=1$ (blue), $n=2$ (red), and $n=3$ (green) for $T_{1u}$ modes (a) $\bar S_1$ [Eq.~\eqref{eq:S1}], (b)
     $\bar S_2$, [Eq.~\eqref{eq:S2}], (c)
     $\bar S_3$, [Eq.~\eqref{eq:S3}],
    and (d) $T_{2u}$ mode $\bar S_4$, [Eq.~\eqref{eq:S4}].   
    The experimental frequencies $\omega_{\bm q=\bm 0, \lambda}$ have been used in Eq.~\eqref{eq:LambdaQE} (Appendix~\ref{app:QEdecomposition}). 
    All modes show linear-in-$k$ EPC around $\Gamma$ in agreement with Eqs.~\eqref{eq:G-sym-T1u}-\eqref{eq:G-sym-T2u}.
    }
    \label{fig:intradfpt}
\end{figure*}

The atomic displacements of the zone-center optical modes in KTO can be given in terms of a complete set of symmetry coordinates~\cite{1967Axe} for the $T_{1u}$ ($\bar S_1, \bar S_2, \bar S_3$) and $T_{2u}$ modes ($\bar S_4$). In terms of atomic displacements $(s^\mathrm{K},s^\mathrm{Ta},s^{\mathrm{O}_x},s^{\mathrm{O}_y},s^{\mathrm{O}_z})$ along $[001]$ they read:
\begin{align}
\label{eq:S1}
    {\bm \bar S_{1}^z}&=\frac{1}{1+\kappa_1}(0,-\kappa_1,1,1,1),  \\
   \label{eq:S2}
     {\bm \bar S_{2}^z}&=\frac{1}{1+\kappa_2}(-\kappa_2,1,1,1,1), \\
   \label{eq:S3}
    {\bm \bar S_3^z}&=\frac{2}{3}(0,0,-\frac{1}{2},-\frac{1}{2},1),\\   
   \label{eq:S4}
    {\bm \bar S_4^z}&=\frac{1}{2}(0,0,1,-1,0)
 \end{align}
 with $\kappa_1=\frac{3M^\mathrm{O}}{M^\mathrm{Ta}}$ and $\kappa_2=\frac{3M^\mathrm{O}+M^\mathrm{Ta}}{M^\mathrm{K}}$. The partners $\bar S_\lambda^x, \bar S_\lambda^y$ polarized along [100] and [010] are found by the corresponding cyclic operations on atom sites $s^{\mathrm{O}_i}$. 
 From their combination, one can define any mode displacement projected onto the desired direction  $\hat{\bm n}$, enforcing the three-fold degeneracy of each mode due to the cubic symmetry. 

The corresponding eigenvectors are written as
\begin{equation}
\label{eq:frometoeta}
    { e}^{\lambda}_{\kappa\alpha} ({\bm q=\bm 0})\equiv \sqrt{M^{\kappa}}\eta_{\kappa\alpha}^{\lambda}=\frac{\sqrt{M^{\kappa}}}{\sqrt{\mu_{\lambda}}}\bar S_\lambda^{\kappa\alpha} \hat{n}_{\alpha}
\end{equation}
with reduced masses 
 \begin{align}
     (\mu_{1})^{-1} & = (M^\mathrm{Ta})^{-1}+(3M^\mathrm{O})^{-1}\\
     (\mu_{2})^{-1} & = (M^\mathrm{K})^{-1}+(M^\mathrm{Ta}+3M^\mathrm{O})^{-1}\\
     (\mu_{3})^{-1} & = (M^\mathrm{O})^{-1}+(2M^\mathrm{O})^{-1}\\
     (\mu_{4})^{-1} & =2(M^\mathrm{O})^{-1}    
 \end{align}
and Eq. \eqref{eq:LambdaQE0} becomes
\begin{align}
\Lambda_{nm,\lambda}(\bm k)=\sum_{\kappa\alpha}\sqrt{\frac{\hbar} {2 \omega_{0\lambda} \mu_{\lambda}}}\gamma^{\kappa\alpha}_{nm} (\bm k) \bar S_\lambda^{\kappa\alpha} \hat{n}_{\alpha}.
\label{eq:LambdaQE}
\end{align}
While there is a single possible mode of $T_{2u}$ symmetry ($\overline{S}_4$), the eigenvectors of the three $T_{1u}$ modes in KTO can in principle be any linear combination of $\overline{S}_1, \overline{S}_2$ and $\overline{S}_3$ modes. Experimentally, however, it was established the eigenvectors ${\bm e}^{\lambda}_{\kappa}(\bm q= \bm 0)$ of the three zone center  $T_{1u}$ modes to be almost pure $\overline{S}_1, \overline{S}_2$ and $\overline{S}_3$ modes, from lowest to highest frequency~\cite{Harada1970,Vogt1988,vogt1995} (in Appendix~\ref{app:QEdecomposition} we show that this is the case also from first-principles calculations). 
That is, the soft TO mode is a nearly `pure' $\bar{S}_1$ mode, with a vibration of the Ta against the oxygen cage. In order to avoid the subtleties involved in a meaningful computation of phonon frequencies and eigenvectors from \emph{ab initio} methods (in particular for modes with a strong temperature dependence, such as the soft TO mode), we will use the experimental frequencies and eigenvectors Eq.\eqref{eq:frometoeta} when computing the EPC Eq.\eqref{eq:LambdaQE}. In particular, in the following, we will adopt at $\bm q=\bm 0$, $\omega_{01}=2.5$ meV,  $\omega_{02}=24.7$ meV, $\omega_{03}=67.7$ meV for the $\overline{S}_1, \overline{S}_2$ and $\overline{S}_3$ modes respectively and corresponding to the TO modes measured in Refs.~\cite{Vogt1988, vogt1995}. 

We compute Eq.~\eqref{eq:LambdaQE} for KTO at the zone center $\bm q=\bm 0$, with the computational parameters and procedures listed in Appendix \ref{app:compdet}. 
Since the zone center phonons of KTO are odd-parity modes, we do indeed find,  in agreement with the symmetry arguments leading to Eq.~\eqref{eq:G-sym-T1u} and Eq.~\eqref{eq:G-sym-T2u}, that for intraband processes and $\bm q=\bm 0$, the pseudospin-independent processes in Eq.~\eqref{eq:Lambda} vanish, $g_{nn,\lambda}(\bm k)\sigma_0=0$.

We can take the 2$\times$2 EPC
matrix for each band to be diagonal by transforming to an appropriate basis at each $\bm k$ point. Then the EPC takes the following pseudospin-dependent form,
\begin{equation}
    \Lambda_{nn,\lambda}(\bm k)=\bm{G}_{nn,\lambda}(\bm k)\cdot \bm{\sigma}=\mc{G}_{n,\lambda}(\bm k) \sigma_3 .  \label{eq:QE-intra}
\end{equation}
Here, for ease of notation, we are using the same symbol 
$ \Lambda_{nn,\lambda}(\bm k)$ for the matrix written in the original basis and the matrix in the diagonal basis.
Note that the invariant in Eq.~\eqref{eq:avrg} is defined so that, for the intraband $n=m$ and $q=0$ case, it matches $\mathcal{G}_{n,\lambda}(\bm k)$ in Eq.\eqref{eq:QE-intra}.

The resulting intraband EPC coupling $\mc{G}_{n,\lambda}(\bm k)$ for the zone-center modes $\bar S_{1,2,3,4}$  polarized along  $\hat{\bm n}=[001]$, and for the perpendicular $M-\Gamma-X$ electronic momentum $\bm k$ path are shown in Fig.~\ref{fig:intradfpt}. 
In agreement with previous frozen phonon computations~\cite{gastiasoro2022theory,venditti2023,gastiasoro2023}, the EPC is linear-in-$k$ for $\bm k\rightarrow 0$.   We have checked that these results are in agreement with frozen phonon computations with 
a root mean square error  $\lesssim 0.01$ meV (See Appendix~\ref{app:frph}). As seen, the obtained coupling to the soft $\bar S_1$ mode is indeed by far the largest.
Also in agreement with Eqs.~\eqref{eq:G-sym-T1u} and \eqref{eq:G-sym-T2u} for band $n=3$,  the splitting has the same initial slope as a function of the distance from the origin. In contrast, bands $n=1$ and $n=2$ show a strong anisotropy of the slope ([100] vs [110]), deviating from the conventional isotropic Rashba splitting.  

\begin{figure}
\includegraphics[width=\columnwidth]{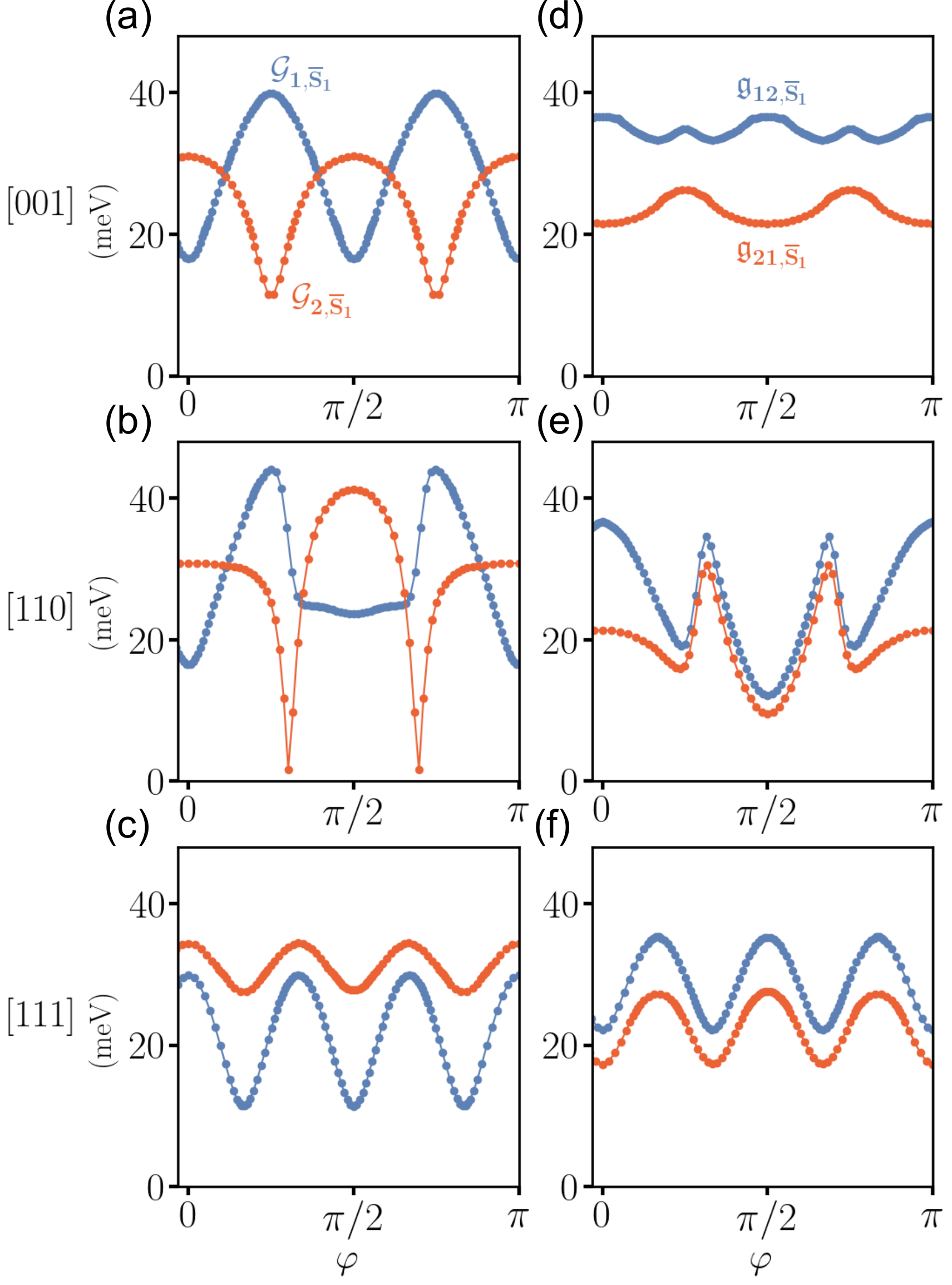}
    \caption{
    \emph{Orientation dependence of the $\bm q=\bm 0$ EPC to the $\bar S_1$ mode}. Intraband $\mc{G}_{n,\bar S_1} (\bm k,\bm q=\bm 0)$, along the FS of band $n=1$ (blue) and the FS of band $n=2$ (red) vs azimuthal angle $\varphi$ (see Fig.~\ref{fig:FS}(b)) in the plane $(k_1,k_2,0)$ perpendicular to (a) [111], (b) [110] and (c) [001]. The Fermi energy is $E_{F,1}=40$ meV and the FS of the planes are shown in Figs.~\ref{fig:FS}(b)-(c). The phonon polarization is perpendicular to the plane examined. 
    The interband matrix elements $\mathfrak{g}_{nm,\bar S_1} (\bm{k},\bm q=\bm 0)$ with $m\neq n$ for the same orientations are shown in panels (d)-(f).
    For interband, we show the matrix element for the scattering of a fermion from the FS of one band to the other band at the same $\bm k$ point (since $\bm q=\bm 0$), which is outside the FS.
    }
    \label{fig:epc-phi}
\end{figure}

Because of experimental constraints, the bands $n=1$ and $n=2$ are precisely those becoming populated upon doping, as shown in Fig.\ref{fig:FS}(a) for two illustrative Fermi energies. We are not aware of experiments in which the $n=3$ band of KTO is populated, either in interfaces or in bulk. Following the strong directional dependence of the superconducting $T_c$ in KTO-based heterostructures, and aiming at a deeper analysis of EPC anisotropies of bands $n=1$ and $n=2$ (coalescing to $j=3/2$ quartet at $\Gamma$), we now compute the EPC for three relevant FS cuts, in the planes perpendicular to cubic direction [001], [110], and [111] at $E_{F,1}=40\,$meV, shown in Figs~\ref{fig:FS}(b)-(d). Here, we assume that the soft eigenmode at the surface is still mostly $\bar S_{1}$~\cite{chu2024}, and we take the phonon polarization perpendicular to the examined surface.

Figs.~\ref{fig:epc-phi}(a)-(c) show the intraband EPC for mode $\bar S_1$ computed along those three FS orientations, respectively, as the azimuthal angle $\varphi$ varies along the FS of each band.
As seen, the intraband EPC has comparable strength for the three different orientations. This implies that without more detailed computations, no clear clues emerge at this level for the orientational dependence of the $T_c$ on surfaces. 

In addition, the EPC shows a rather strong azimuthal angular dependence, also seen in Fig~\ref{fig:intradfpt}(a) and previous frozen phonon computations~\cite{venditti2023}. 
We anticipate that, when compared to a three-band model (Section~\ref{sec:model}), it will be necessary to consider Rashba-type-II processes (spin-non-conserving) to capture the correct angular dependence of the EPC to the Slater $\bar S_1$ mode. 
This was marginally discussed already in Ref.~\cite{venditti2023}, where a frozen phonon computation along high symmetry lines was already showing the different anisotropic behavior of the Rashba splitting for the various modes. 

A great advantage of DFPT calculations is having access to the interband EPC matrix elements, unlike the frozen phonon method. 
In the interband channel, the pseudospin independent terms $g_{nm,\lambda}(\bm k,\bm q=\bm 0)\sigma_0$ with $n\neq m$, can now also be finite. We indeed find the interband matrices to be of the general form given by Eq.~\eqref{eq:Lambda}.
We show in Figs.~\ref{fig:epc-phi}(d)-(f) the corresponding invariants using Eq.~\eqref{eq:avrg} for the lowest bands, $\mathfrak{g}_{12}(\bm k)$ and $\mathfrak{g}_{21}(\bm k)$, along the FS for the same three orientations $[001]$, $[110]$ and $[111]$, respectively. As seen, these interband matrix elements are of the same order of magnitude as the intraband matrix elements in panels (a)-(c) in the same figure. Since we are considering ${\bm q}=\bm 0$ processes, we show the matrix element for the scattering of a fermion from the FS path of one band to the other band at the same $\bm k$ point, which will be outside the FS. Because of this, the two matrix elements are different ($nm=12$ in blue, and $nm=21$ in red).

For completeness, we present the same results for the other two $T_{1u}$ modes, $\bar S_{2}$ and $\bar S_{3}$, and the $T_{2u}$ mode $\bar S_4$ in Appendix \ref{app:DFPT_S2_S3}. Also in this case, despite the strong angular dependence, there is no evident dominance of coupling in any direction. 

\subsection{Finite $\mathit{{\bf q}}$ modes}

\begin{figure}
    \centering
    \includegraphics[width=1\linewidth]{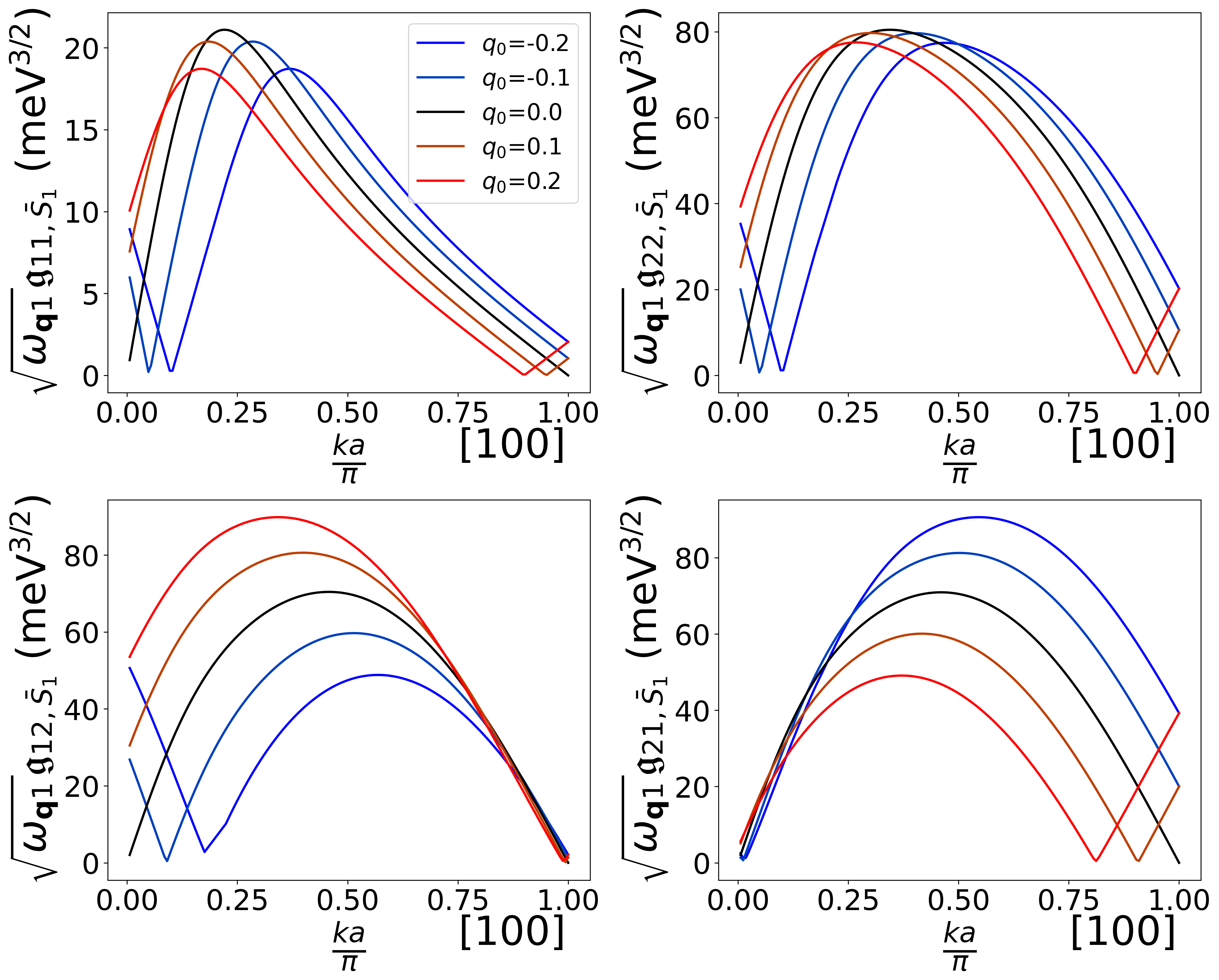}
    \caption{
    Intraband  $\sqrt{\omega_{\bm q 1}}\mathfrak{g}_{11/22,\bar{S}_{1}}(\bm k+\bm q/2,\bm q)$ and interband matrix elements $\sqrt{\omega_{{\bm q}1} }\mathfrak{g}_{12, 1}(\bm k+\bm q/2,\bm q)$ using invariant Eq.~$\eqref{eq:avrg}$. The polarization of the $\bar S_1$ mode is along [001] with $\bm k$ along [100] and $\bm{q}=(q_0,0,0)$ with $q_0=0,\pm 0.1, \pm0.2$ (in units of $\pi /a$). 
    }
    \label{fig:EPC_combined}
\end{figure}

Within the DFPT framework, it is relatively easy to access electron-phonon matrix elements also at finite ${\bm q}$. However, differently from the zone-center case, the momentum dependence of the phonon frequency $\omega_{{\bm q}\lambda}$ and of the phonon eigenvector $e^{\lambda}_{\kappa\alpha}(\bm q)$ can in principle have a significant impact on the small-$\bm q$ behavior of  $\Lambda_{nm,\lambda}(\bm k,\bm q)$. Since the experimental determination of the eigenvector of the KTO soft mode at finite $\bm q$ has not been done, we concentrate on the ${\bm 
q}$ dependence of the matrix elements $\gamma^{\kappa\alpha}_{nm} (\bm k,\bm q)$, by computing $\sqrt{\omega_{\bm q \lambda}}\Lambda_{nm,\lambda}(\bm k,\bm q)$ approximating $e^{\lambda}_{\kappa\alpha}(\bm q)$ with the zone-center eigendisplacement of Eq.~\eqref{eq:frometoeta}. We show in Fig.~\ref{fig:EPC_combined} these intraband $\sqrt{\omega_{\bm q1}}\mathfrak{g}_{11/22,\bar{S}_1}(\bm k+\bm q/2,\bm q)$ and interband $\sqrt{\omega_{\bm q1}}\mathfrak{g}_{12/21,\bar{S}_1}(\bm k+\bm q/2,\bm q)$ terms for a $\bar{S}_1$ mode polarized along [001], $\bm k$ along the perpendicular [100] direction,  and $\bm{q}=(q_0,0,0)\parallel {\bm k}$ with $q_0=0,\pm 0.1, \pm0.2$ (in units of $\pi /a$). 

In general, the presence of a finite $\bm q$ introduces both the dependence of $\gamma^{\kappa\alpha}_{nm}(\bm k, \bm q)$ upon the angle between $\bm k+ \bm q/2$ and $\bm k- \bm q/2$, and on their moduli $|\bm k+ \bm q/2|$,$|\bm k- \bm q/2|$. These dependencies shift the minima of $\mathfrak{g}_{nn,\lambda}(\bm k +\bm q/2,\bm q)$ away from $\bm k=\bm 0$ to $\bm k=- \bm q/2$ at finite $\bm q$, signaling the presence of trigonometric factors. Equivalently, the minima of $\mathfrak{g}_{nn,\lambda}(\bm k,\bm q)$ always appear at $\bm k=0$ at finite $\bm q$. This is consistent with the fact that in the presence of time-reversal symmetry $|\mathfrak{g}_{nn,\lambda}(\bm k,\bm q)|=|\mathfrak{g}_{nn,\lambda}(-\bm k,\bm q)|$ must hold.
Additional dependencies on $|\bm q|$ are instead much milder up to the largest ${\bm q}$ analyzed [Fig. \ref{fig:EPC_combined}]. This is probably a general feature of the EPC in this system, but its systematic study will be left for future investigations. 

Finally, we note that finite $\bm q$ calculations are continuous to $\bm q=\bm 0$ ones, signaling the absence of long-range components in the electron-phonon coupling of the analyzed modes, as expected for TO phonons \cite{PhysRevLett.129.185902,PhysRevB.107.094308,PhysRevB.110.094306}.
Given the weak $|q|$-dependence found for small doping, it is reasonable to take the $\bm q=\bm 0$ results as representative of the interband electron-phonon coupling. 

\section{Three-band model of dynamic Rashba-like coupling
}

\label{sec:model}
In the following, we show how the EPC to polar modes in Eq.~\eqref{eq:Hepc} emerges when considering induced electronic hopping processes in real space in a three-band relativistic model with the $t_{2g}$ orbitals of Ta~\cite{gastiasoro2023}. 
Indeed, in the presence of a polar distortion, new terms appear in the electronic Hamiltonian. Identifying the dominant ones can help to understand the momentum structure of the EPC (Fig~\ref{fig:epc-phi}), determine the relevance of spin-conserving (type I) and spin-non-conserving (type II) processes, and pinpoint the microscopic processes that lead to the symmetry form in Eq.~\eqref{eq:G-sym-T1u}.

We now introduce a three-band model (spin-orbit coupled $t_{2g}$ orbitals) to describe the coupling between the electrons and polar phonons in this system. In the presence of a polar mode $\lambda$ 
to linear order in the distortion, there are new induced hopping terms $t_{\mu\nu,j,\lambda}(\bm k, \bm q)$. The EPC Hamiltonian reads, 
\begin{flalign}    
\label{eq:Hu}    &\mathcal{H}_\mathrm{EPC}=\sum_{\bm k\bm q,\mu\nu j} \varphi^\dagger_\mu(\bm k+\frac{\bm q}{2}) t_{\mu\nu,j,\lambda}(\bm k,\bm q)\sigma_j\varphi_\nu(\bm k-\frac{\bm q}{2})\mc{A}_{\bm q,\lambda}&
\end{flalign}
with Pauli matrices $\sigma_j$ representing electronic spin-independent ($j=0$) and spin-dependent ($j=1,2,3$) hopping processes between orbitals $\mu$ and $\nu$ on neighboring Ta sites. 

To identify the most important induced hopping terms, we performed a $\bm q=\bm 0$ frozen phonon computation together with a Wannier projection of the electronic structure with and without the lattice distortion for the three polar modes. We employed maximally localized Wannier functions \cite{wannier90_RMP} with a polar phonon oriented along $[001]$.
The Wannier projection was restricted to the $t_{2g}$ manifold, so all the indirect processes via different orbitals are effectively incorporated in the projection. For further details, see Appendix~\ref{app:wannier}.  

\begin{figure}
    \includegraphics[width=\columnwidth]{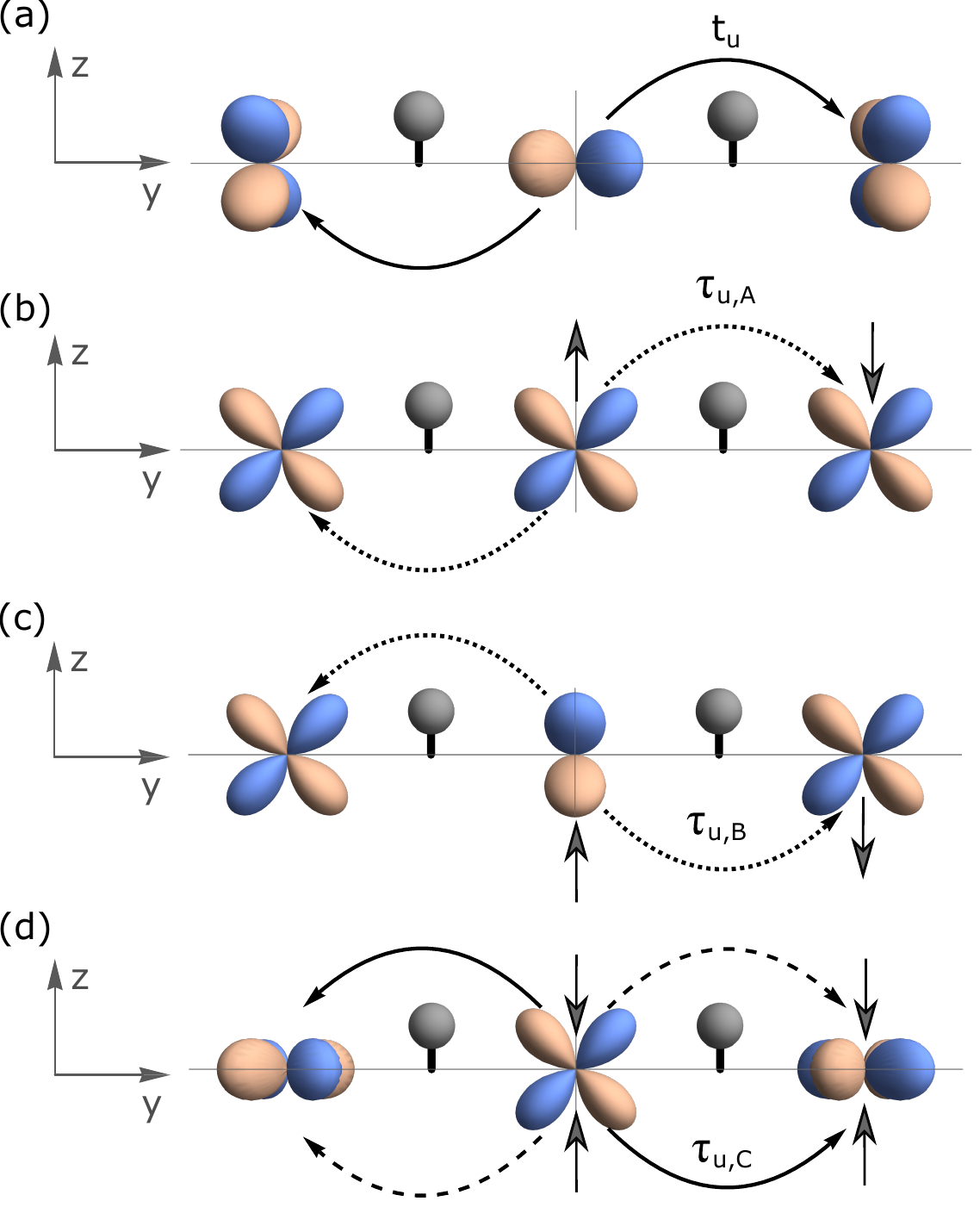}
    \caption{\emph{Schematic induced odd-parity hopping terms in a structure with a polar phonon along $\hat{\bm n}=\hat{\bm z}$}. The phonon is represented by the polar displacement of the oxygen atoms (gray spheres) along $\hat{z}$, with induced hopping channels between NN $t_{2g}$ orbitals along the $\hat{y}$-bond. (a) Spin-independent interorbital $t_{yz,0}(\bm k)\sigma_0$ [Eq.~\eqref{eq:tu}], (b) spin-flip intraorbital $t_{xx,1}(\bm k)\sigma_1$ [Eq.~\eqref{eq:tau1}], (c) spin-flip interorbital $t_{xy,2}(\bm k)\sigma_2$[Eq.~\eqref{eq:tau2}], and (d) spin-dependent interorbital $t_{zx,3}(\bm k)\sigma_3$ [Eq.~\eqref{eq:tau3}]. The spin of the orbitals in the relativistic processes is represented by gray arrows, only shown for $y>0$ for clarity. Notice that the flipping or not of the spin depends on the spin state of the electron considered. We schematically illustrate the processes for spins polarized along the $z$ direction.
    }
    \label{fig:tu}
\end{figure}

We find strong differences between the odd-parity induced terms by each polar mode $\bar S_\lambda$ in KTO. The following four hopping and their symmetry-related terms between nearest-neighbor (NN) Ta atoms dominate the response,
\begin{flalign}
\label{eq:tu}
    t_{yz,0}(\bm k)&=-2it_u \sin k_y,  & t_{zx,0}(\bm k)&=2it_u \sin k_x, \\
    \label{eq:tau1}
    t_{xx,1}(\bm k)&=2\tau_{u,A} \sin k_y,  & t_{yy,2}(\bm k)&=-2\tau_{u,A} \sin k_x,\\
     \label{eq:tau2}
     t_{xy,2}(\bm k)&=-2\tau_{u,B} \sin k_y, & t_{xy,1}(\bm k)&=2\tau_{u,B} \sin k_x,\\
      t_{zx,3}(\bm k)&=-2\tau_{u,C} \sin k_y,  & t_{yz,3}(\bm k)&=2\tau_{u,C} \sin k_x.  
   \label{eq:tau3}
\end{flalign}
Note that we dropped the ${\bm q}$ dependence and $\lambda$ subscript to simplify the notation. 
The amplitude $t_u$ in Eq.~\eqref{eq:tu} corresponds to the previously studied spin-independent (type I) process~\cite{gastiasoro2022theory,venditti2023}, and it is shown in Fig~\ref{fig:tu}(a). The other three terms in Eqs.~\eqref{eq:tau1}-\eqref{eq:tau3} correspond to spin-dependent (type II) processes.
The induced term in Eq.~\eqref{eq:tau1} is shown in Fig~\ref{fig:tu}(b) and describes 
a spin-non-conserving intra-orbital hopping with amplitude $\tau_{u,A}$. 
Fig~\ref{fig:tu}(c) describes also a spin-non-conserving term, involving two different $t_{2g}$ orbitals instead [Eq.\eqref{eq:tau2}]. The last type II term [Eq.~\eqref{eq:tau3}] in Fig~\ref{fig:tu} (d) is an interorbital hopping with spin-dependent amplitude $\tau_{u,C}$, opposite for each spin projection along $z$. The analogous terms for a polar phonon along $[100]$ and $[010]$ can be straightforwardly derived using a permutation of the orbitals.

Notice that Eq.~\eqref{eq:Hepc} and Eq.~\eqref{eq:Hu} represent the same electron phonon interaction, the former in the band basis and the latter in the orbital basis. The matrix elements  $\Lambda_{nm}$ and $t_{\mu\nu,j}$ are related by a unitary transformation and are derivatives of an energy with respect to atomic displacements multiplied by the harmonic oscillator characteristic length 
$l_{\bm q\lambda}\equiv \sqrt{{\hbar} /(2 \omega_{\bm q\lambda} \mu_\lambda})$  [c.f. Eqs.~\eqref{eq:gamma},~\eqref{eq:LambdaQE}] and therefore have units of energy. 
Likewise, the induced hopping parameters have units of energy, and their value depends on the choice of the phonon frequency, while their ratios reported below do not.

Table~\ref{tab:tu} shows the ratio of the induced type II hopping amplitudes $\tau_{u,i}$ to the type I amplitude $t_u$ for the three polar modes $\bar S_\lambda$ in KTO [Eqs.\eqref{eq:S1}-\eqref{eq:S3}] obtained from the Wannier projection. While the ratios are quite small for the $\bar S_2$ mode, they become intermediate for $\bar S_3$, and more importantly, of the order of one for $\bar S_1$. This indicates that the spin-non-conserving terms in the EPC vertex for the Slater mode $\bar S_1$ are as relevant as the spin-conserving inter-orbital $t_u$ term. This is consistent with the much stronger intraband interaction for all bands in Fig.~\ref{fig:intradfpt}(a) compared to panels (b) and (c), which cannot be reconciled solely with the mode-dependent $t_u$ magnitudes. 
The magnitude of $t_u$ is discussed in the next subsection, as it was determined by a different method. 

\begin{table}[]
    \centering
    \begin{tabular}{c|c c c}
                         & $\bar{S}_1$ & $\bar{S}_2$ & $\bar{S}_3$ \\ \hline
       $\tau_{u,A}/ t_u$  & 1.08 & 0.18 & -0.35 \\
       $\tau_{u,B}/ t_u$  & 0.1 & -0.01 & 0.31 \\
       $\tau_{u,C}/ t_u$  & 0.44 & 0.06 & -0.33 \\
       $t_u$ [meV]        & 48.8 & 4.3  & 8.3
    \end{tabular}
    \caption{
    Hopping terms induced by the three polar modes $\bar{S}_i$ in KTO [Eqs.~\eqref{eq:S1}\eqref{eq:S3}]. We report the  ratio between spin-non-conserving (type II) (Eqs.~\eqref{eq:tau1}-\eqref{eq:tau3}) and spin-conserving $t_u$ (type I)  terms (Eq.~\eqref{eq:tu}), obtained from localized Wannier functions.   The $t_u$ hopping is obtained from a linear-in-$k$ fit to the EPC obtained in a non-relativistic DFPT calculation (see Appendix~\ref{app:tu}) and using the experimental phonon frequencies listed in Section~\ref{sec:dfptdef}.}
    \label{tab:tu}
\end{table}

We use now the three-band model of $t_{2g}$ orbitals with SOC [Eq.~\eqref{eq:H0} and Eq.\eqref{eq:Hu}] to show how the spin-non-conserving terms determine the momentum structure and strength of the polar EPC. 

\subsection{Rashba type I processes}
\label{sec:nosoc}

Let us start by neglecting SOC ($\xi=0$) in Eq.~\eqref{eq:H0} and doing the analysis for the high-symmetry direction $\hat{\bm{k}}=[100]$ to obtain analytical expressions. The tight-binding model $\mathcal H_0=\sum_{\bm k} h_0(\bm k)$ of the three spin-$1/2$ $t_{2g}$ orbitals takes the form
\begin{equation}
\label{eq:H0nosoc100}
    h_0(\xi=0,k_x)= \sum_\mu\varphi_\mu^\dagger(k_x) E_\mu(k_x)\sigma_0\varphi_\mu(k_x) 
\end{equation}
with energy dispersions
\begin{align}
     E_x(k)&=2t_2(1-\cos k)\\
    E_y(k)&=E_z(k)=2(t_1+2t_3)(1-\cos k)
\end{align}
Again, $\sigma_0$ is the identity matrix acting on the electronic spin-$\frac{1}{2}$ sector of the  $t_{2g}$ orbitals. Hence, along this $\hat{\bm k}=[100]$ direction, we have a doubly degenerate band with orbital character $\mu=x$ and a fourfold-degenerate band with orbital character $\mu=y,z$. 
 The same holds for the equivalent $\hat{\bm{k}}=[010]$ direction, but interchanging the $x$ and $y$ orbital character.

Without SOC, a polar phonon can only induce the interorbital spin-independent processes $t_u$ [Eq.~\eqref{eq:tu}].
In the electronic basis $\varphi_\mu$ in Eq.~\eqref{eq:H0nosoc100}, these induced hopping elements for a phonon polarized along $[001]$ become pure \emph{interband} terms, mixing the orbitals $\mu=x,y$ in the lower band with the $\mu=z$ member of the upper band manifold:
\begin{flalign}
\label{eq:Hukxnosoc}
    &\mc{H}_\mathrm{EPC}=\sum_{\bm k}2it_{u}\left[\varphi_x^\dagger(\bm k) \sin(k_x)\sigma_0\varphi_z(\bm k)\right.&\\\nonumber
    &\left.\qquad\quad+\varphi_y^\dagger(\bm k)\sin(k_y)\sigma_0\varphi_z(\bm k)\right]\mc{A}_{q=0}+\mathrm{h.c.}\ \ \ \ \mathrm{(Type~I)}&
\end{flalign}
Note that the double spin degeneracy is preserved (the spin sectors up and down are still completely decoupled), i.e., the polar term only acts in the band sector, as an interband spin-conserving coupling.

The  $t_u$ parameter could in principle be directly obtained from the relativistic Wannier projection discussed above, which includes this nearest-neighbour hopping amplitude. Instead, we can obtain an \emph{effective} $t_u$ parameter taking into account further Ta neighbors, beyond NN, in the spirit of tight-binding models (for details see Appendix~\ref{app:tu}).  We perform a non-relativistic DFPT computation for a polar mode $\bar S_\lambda$. In this case, the second term in  Eq.~\eqref{eq:Lambda} vanishes and Eq.~\eqref{eq:Hukxnosoc} contributes only to the first term, i.e., $\Lambda_{nm,\lambda}(\bm k)= g_{nm,\lambda}(\bm k) \sigma_0$ for $n\neq m$. 
By fitting the DFPT results with the model [Eq. \eqref{eq:H0nosoc100} and Eq. \eqref{eq:Hukxnosoc}], we obtain the effective $t_u$ values in Table~\ref{tab:tu}. 

The $\bar S_2$ mode shows the smallest EPC amplitude, more so when considering $\sqrt{\omega_{0\lambda}} t_u$. This may not be so surprising, since from Eqs.~\eqref{eq:S1}-\eqref{eq:S3}, one can see that this is the only mode preserving the Ta-O bond.
We have also checked that in the presence of SOC the $t_u$ parameters from Wannier do not change significantly, the change for the soft $\bar S_1$ mode being less than 1$\%$.  

Switching on the SOC, we can now project the spin-conserving EPC model Eq.\eqref{eq:Hukxnosoc} into the relativistic electronic spinors $\psi_n^\dagger(\bm k) $ in Eq.~\eqref{eq:H0soc}. 
It is instructive to analyze first the problem in the small $k$ limit, neglecting quadratic and higher order in $k$ corrections, where the electronic basis Eq.~\eqref{eq:fi2psi} diagonalizes $\mc{H}_0$.
Projecting the matrix elements of the type-I EPC model Eq.~\eqref{eq:Hukxnosoc} induced by the parity-breaking phonon to the  SOC basis in Eq.~\eqref{eq:fi2psi} we obtain,
\begin{flalign}
    \label{eq:Hutsoc}
&\mc{H}_\mathrm{EPC}=\left[\sum_n\psi_n^\dagger \vartheta_n(k_x 
\sigma_2-k_y\sigma_1)\psi_n\right.&\\ \nonumber
&\left.+\psi_1^\dagger \vartheta_{12}(k_x 
\sigma_2+k_y\sigma_1)\psi_2+\psi_1^\dagger \vartheta_{13}(k_x 
\sigma_2-k_y\sigma_1)\psi_3 \right.&  \\ \nonumber
&\left.+\psi_2^\dagger \vartheta_{23}(k_x 
\sigma_2+k_y\sigma_1)\psi_3\right]\mc{A}_{q=0}+\mathrm{h.c.}&
\end{flalign}
with the following intraband and interband coupling terms
\begin{flalign}
\label{eq:intra1tu}
    &\vartheta_1=-\vartheta_3=-\frac{4}{3}t_u&\\
    \label{eq:intra2tu}
    &\vartheta_2=0& \\
    &\vartheta_{12}=-\sqrt{6}\vartheta_{13}=-\sqrt{2}\vartheta_{23}=\frac{2}{\sqrt{3}}t_u\label{eq:intertu}&
\end{flalign}
For band 3, 
the polar phonon induces a pseudospin dependent intraband Rashba-like  term $\vartheta_3$, recovering the expression in Eq.~\eqref{eq:G-sym-T1u} derived from symmetry arguments for a $T_{1u}$ mode, with coupling constant $\frac{4}{3}t_u$. This leads to the conventional isotropic Rashba splitting in the presence of a static lattice distortion. 

A similar result is found for the intraband term of band 1,
but, in addition, the phonon also induces pseudospin-dependent $\vartheta_{nm}$ interband matrix elements, with a Rashba-like coupling $\vartheta_{13}$ and a Dresselhaus-like coupling for $\vartheta_{12}$ and $\vartheta_{23}$. 
Notice that because of the degeneracy of bands 1 and 2 in this $k$ limit, what is called ``intraband" and ``interband" is basis dependent, and the 4$\times$4 matrix with the interband interaction has to be taken into account to obtain the splitting.

By construction, we considered only type I processes in Eq.~\eqref{eq:Hukxnosoc}. Therefore, all EPC terms in Eqs.\eqref{eq:intra1tu}-\eqref{eq:intertu} are proportional to the spin-conserving hopping $t_u$ involving inter-orbital amplitudes [Eq.\eqref{eq:tu}].  In this case,  SOC is only needed to modify the electronic structure (from non-relativistic to relativistic), but it is not critical for the existence of a finite electron-phonon vertex~\cite{volkov2020,gastiasoro2022theory,chaudhary2023,gastiasoro2023}. 
In the absence of SOC, the electron-phonon vertex does not vanish but simply becomes a spin-conserving interband term, as in Eq.~\eqref{eq:Hukxnosoc}.

\begin{figure}
    \includegraphics[width=\columnwidth]{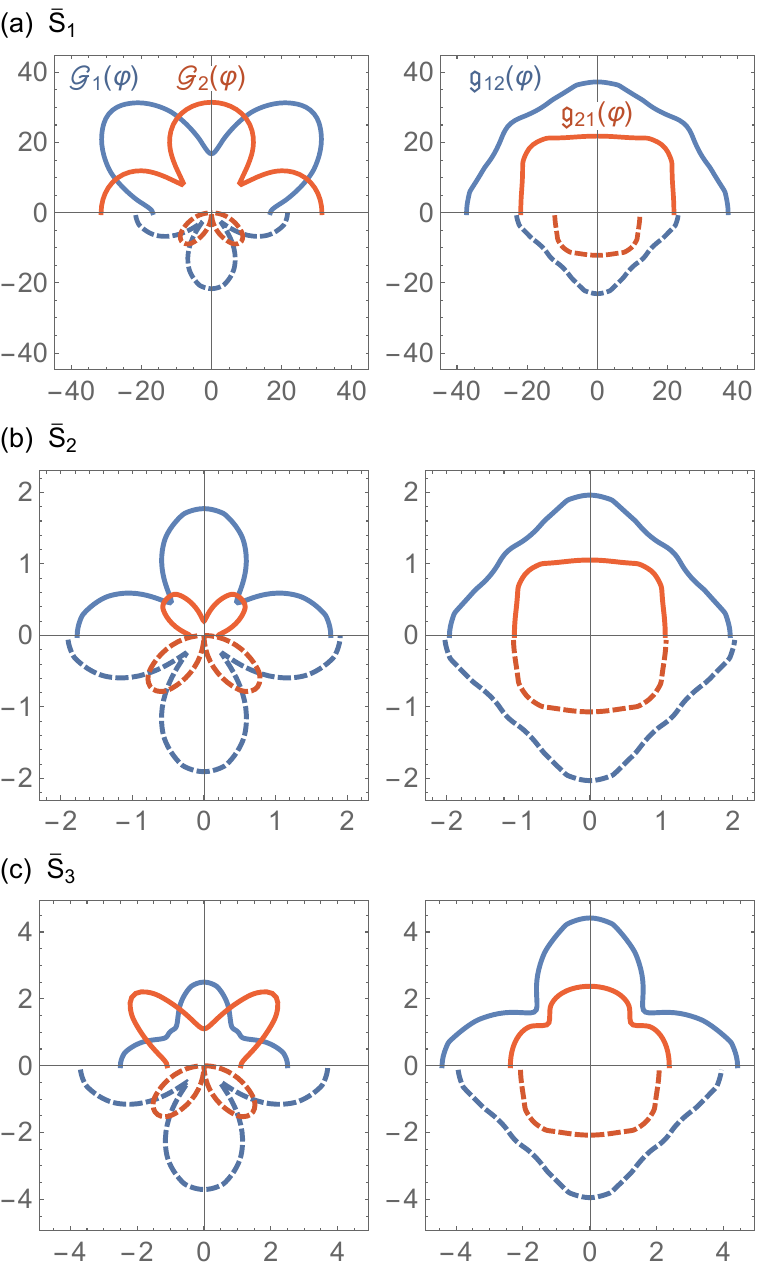}
    \caption{\emph{Type I EPC model ($\tau_{u,i}=0$) vs DFPT}. Polar plot of the intraband  $\mathcal{G}_{n,\lambda}(k_F\varphi)$ (left panels) and interband $\mathfrak{g}_{nm,\lambda}(k_F\varphi)$ (right panels) in meV, for a $\bm q = \bm 0$ mode $\lambda$ with polar axis along $[001]$, along the FS of band $n=1$ (blue) and $n=2$ (red) in the perpendicular $k_x-k_y$ plane for $E_{F,1}=40$ meV. In each panel the EPC is computed both by DFPT (for $0<\varphi<\pi$, full lines) and by the three-band model in Eq.~\eqref{eq:H0} and Eq.\eqref{eq:Hu} with $t_u$ from Table~\ref{tab:tu} (for $-\pi<\varphi<0$, dashed lines). (a) $\lambda=\bar{S}_1$ (equivalent to Fig.~\ref{fig:epc-phi}(e)-(f)),  (b) $\lambda=\bar{S}_2$, and  (c) $\lambda=\bar{S}_3$. 
   }
    \label{fig:qe-vs-tb-tu}
\end{figure}

From the analytically derived EPC expressions in Eqs.\eqref{eq:intra1tu}-\eqref{eq:intertu} one can already see that in general the resulting interband $\vartheta_{nm}$ and intraband $\vartheta_{n}$ couplings are of the same order of magnitude ($\sim t_u$), in agreement with the \emph{ab initio} results in Fig.~\ref{fig:epc-phi}. One can show analytically, {going to quadratic order in $k$,} that the type I EPC model also displays a vanishing EPC for band $n=2$ for $k_x=0$ and $k_y=0$~\cite{venditti2023}, which is certainly not the case for the \emph{ab initio} EPC, as can be read from Fig.~\ref{fig:intradfpt}, where the $n=2$ band shows, at small $k$, a sizable linear-in-$k$ intraband EPC along $k_x$ for all polar modes.

Interestingly, as we shall see, this model yields an anisotropic Rashba splitting, which is a necessary condition to explain the DFT results.
However, the obtained angular dependence is qualitatively different for some of the modes. 
To illustrate more clearly the problem with neglecting type II hopping processes in the EPC, 
Fig.~\ref{fig:qe-vs-tb-tu} compares the type I model using Eq. \eqref{eq:H0} and Eq. \eqref{eq:Hukxnosoc} (dashed lines in all panels) with relativistic DFPT computations (full lines in all panels) along the same FS $\bm k$-path. That is, projecting $\mathcal{H}_\mathrm{EPC}$ onto the relativistic electronic basis that diagonalizes $\mathcal{H}_0$ for each $\bm{k}$ point along the FS of each electronic band $n$, we get the corresponding EPC matrix $\Lambda_{nm,\lambda}(\bm k)$ [Eq.~\eqref{eq:Lambda}] for a mode $\lambda$, from which the intraband (left column) and interband (right column) EPC strengths are obtained, and compared to DFPT results. 
The modes are polarized along $\hat{\bm n}\parallel[001]$, and the EPC is shown as a polar plot, i.e., 
along the FS of bands $n=1$ (blue) and $n=2$ (red) in the perpendicular $k_x$-$k_y$ plane, for $E_{F,1}=40$ meV, illustrated in Fig.~\ref{fig:FS}(a).
Note that bands 1 and 2 are no longer degenerate [Fig.~\ref{fig:FS}(b)], unlike in the linear-in-$k$ approximation in Eq.~\eqref{eq:fi2psi}, due to the mixing by the kinetic energy terms in $\mathcal H_0$ at quadratic order in $k$. 

As we already pointed out above and in Ref.~\cite{venditti2023}, the type I model only captures some of the EPC features obtained by \emph{ab initio} in KTO. As seen in Fig.~\ref{fig:qe-vs-tb-tu}(b), both intraband (left) and interband (right) EPCs of mode $\bar S_2$ are beautifully reproduced by this three-band EPC model, both qualitatively and quantitatively. The EPC to the polar mode $\bar S_3$ obtained by the model in Fig.~\ref{fig:qe-vs-tb-tu}(c) shows bigger disagreement with its DFPT counterpart, although the general features are still qualitatively captured. Finally, the model severely fails when describing the EPC to the soft mode $\bar S_1$, as shown in panel (a). In particular, we highlight the substantial qualitative difference found in the intraband contribution, where the maxima and minima appear inverted for both $n=1$ and $n=2$ bands. 

The virtue of the type I EPC model for mode $\bar S_2$ and its shortcomings for mode $\bar S_1$ (and to some extent $\bar S_3$) are not so surprising, if one bears in mind the mode-dependent relevance of the induced non-spin-conserving terms $\tau_{u,i}$ discussed in the beginning of this Section, and listed in Table~\ref{tab:tu}. In the following, we show how these type II terms can considerably affect the EPC structure (and strength) of the three polar modes.

\subsection{Rashba type I and type II processes}
\label{sec:withsoc}
Let us now consider the EPC model by including also the spin-non-conserving terms in Eqs.~\eqref{eq:tau1}-\eqref{eq:tau3}. 
We can again project the matrix elements in Eqs.~\eqref{eq:Hu}-\eqref{eq:tau3} induced by the polar phonon to the SOC electronic basis Eq.~\eqref{eq:fi2psi}, as in Eq.~\eqref{eq:Hutsoc}. The corresponding intraband and interband induced elements, listed in the following, contain now besides $t_u$, the spin-non-conserving terms $\tau_{u,i}$:
\begin{align}
\label{eq:intra1}
    \vartheta_1&=-\frac{1}{3}\left(4t_u-\tau_{u,A}+2\tau_{u,B}-4\tau_{u,C}\right)\\
    \label{eq:intra2}\vartheta_2&=\left(\tau_{u,A}+2\tau_{u,B}\right) \\
    \vartheta_3&=- \frac{1}{3}\left(4t_u+2\tau_{u,A}-4\tau_{u,B}-4\tau_{u,C}\right)
    \label{eq:intra3}\\
    \label{eq:inter12}
    \vartheta_{12}&=\frac{1}{\sqrt{3}}\left(2t_u+\tau_{u,A}+2\tau_{u,C}\right) \\
    \vartheta_{13}&= -\frac{\sqrt{2}}{3}\left(t_u-\tau_{u,A}+2\tau_{u,B}-\tau_{u,C}\right)\\
    \vartheta_{23}&= -\sqrt{\frac{2}{3}}\left(t_u-\tau_{u,A}+\tau_{u,C}\right).
    \label{eq:inter23}
\end{align}
As seen, the spin dependent $\tau_{u,i}$ amplitudes also lead to Rashba-like EPC terms at small momenta [e.g. Eqs.~\eqref{eq:intra1}-\eqref{eq:intra3}].  Contrary to the type I Rashba EPC arising from the spin-independent $t_u$ amplitude explored before, $\tau_{u,i}$ are proportional to SOC (they must vanish without relativistic effects). We thus call the type of EPC originating from $\tau_{u,i}$ spin-dependent vertex, \emph{Rashba type II} EPC. 

\begin{figure}[th!]
    \includegraphics[width=\columnwidth]{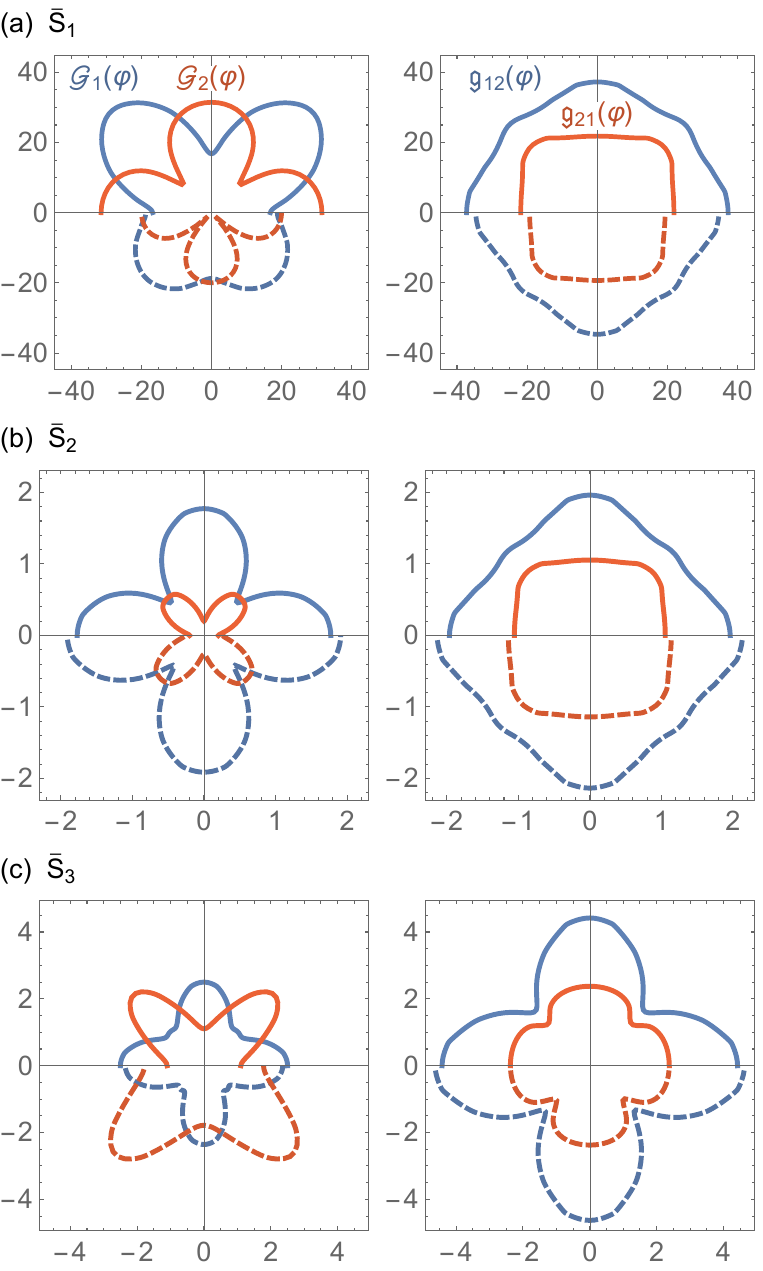}
    \caption{\emph{Type I and type II EPC model ($\tau_{u,i}\neq 0$) vs DFPT}. Polar plot of the intraband  $\mathcal{G}_{n,\lambda}(k_F\varphi)$ (left panels) and interband $\mathfrak{g}_{nm,\lambda}(k_F\varphi)$ (right panels) in meV, for a $\bm q = \bm 0$ mode $\lambda$ with polar axis along $[001]$, along the FS of band $n=1$ (blue) and $n=2$ (red) in the perpendicular $k_x-k_y$ plane for $E_{F,1}=40$ meV. In each panel the EPC is computed both by DFPT (for $0<\varphi<\pi$, full lines) and by the three-band model in Eq.~\eqref{eq:H0} and Eq.\eqref{eq:Hu} with $t_u$ and $\tau_{u,i}$ from Table~\ref{tab:tu} (for $-\pi<\varphi<0$, dashed lines). (a) $\lambda=\bar{S}_1$ (equivalent to Fig.~\ref{fig:epc-phi}(e)-(f)),  (b) $\lambda=\bar{S}_2$, and  (c) $\lambda=\bar{S}_3$.}
    \label{fig:qe-vs-tb}
\end{figure}
It is interesting to note that the induced terms $t_u$, $\tau_{u,A}$, $\tau_{u,B}$ and $\tau_{u,C}$ [Eqs.~\eqref{eq:tu}-\eqref{eq:tau3}] enter the intraband $\vartheta_n$ and interband $\vartheta_{nm}$ EPC of the various bands in distinct ways. For example, the coupling for band $n=2$ [Eq.~\eqref{eq:intra2}] has only spin-non-conserving contributions $\tau_{u,i}$, while bands $n=1,3$ have contributions from the spin-conserving $t_u$ term as well. This already indicates that the relative strength of the induced type II terms $\tau_{u,i}$ with respect to the type I term $t_u$ for a given polar mode determines the final momentum structure of the EPC.   

The dashed lines in all panels of Fig.~\ref{fig:qe-vs-tb} show the intraband (left) and interband (right) EPC by the type I and type II model, i.e. using Eq.~\eqref{eq:H0} and Eq.~\eqref{eq:Hu} with the induced $t_u$ and $\tau_{u,i}$ parameters in Table~\ref{tab:tu} for each $\bar S_i$ polar mode, along the same FS $\bm k$-path used in the DFPT calculations (full lines in the same figure). 
As seen, overall the qualitative match between the three-band model and the \emph{ab initio} results is now very good also for modes $\bar S_1$ and $\bar S_3$, in particular for the interband matrix elements. Also, it is clear that the different weight of the type II induced terms in each polar mode (Table~\ref{tab:tu}) affects both the overall EPC strength (e.g. $\bar S_1$ vs $\bar S_3$, despite having similar $t_u\sqrt{\omega_\lambda}$) as well as the EPC momentum structure (e.g. $\bar S_1$ vs $\bar S_{2,3}$). 

\begin{figure}
    \includegraphics[width=\columnwidth]{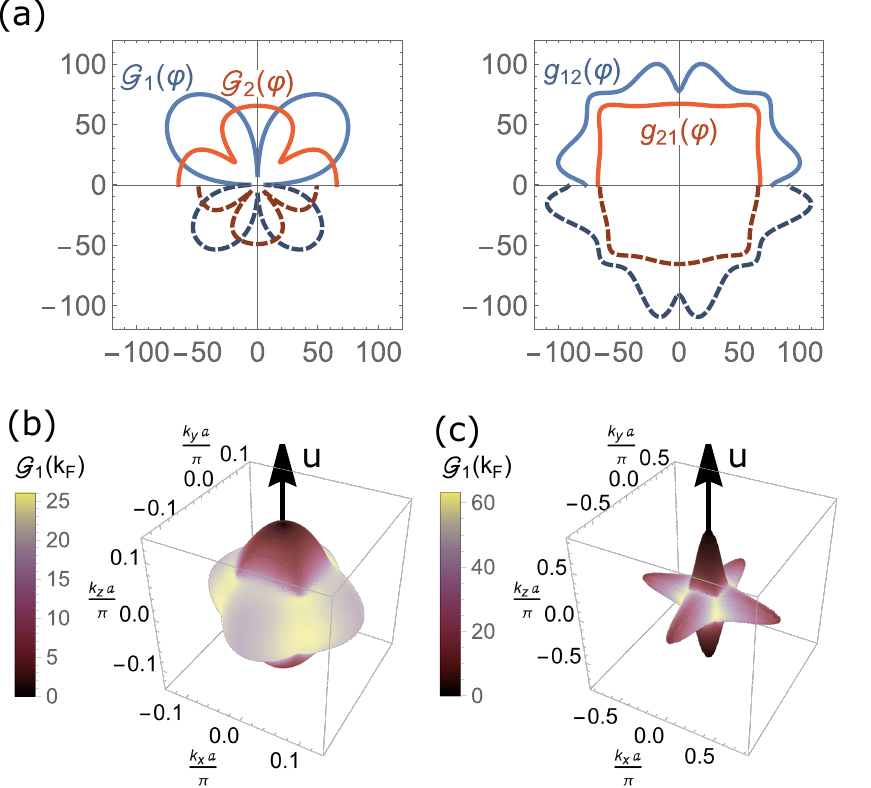}
    \caption{\emph{Three-band model beyond small $k$ and on 3D FSs}. 
    (a) Same as Fig.~\ref{fig:qe-vs-tb}(a) but for higher Fermi energy $E_{F,2}=240$ meV (shown in Fig.\ref{fig:FS}(a)). 
    (b) Intraband EPC $\mathcal{G}_{1,\bar{S}_1}(k_F\hat{\bm k})$ for a $\bar{S}_1$ mode polarized along $[001]$ (black arrow), projected on the 3D FS for $E_{F,1}=40$ meV and (c) $E_{F,2}=240$ meV. Obtained by the three-band model in Eq.~\eqref{eq:H0} and Eq.~\eqref{eq:Hu} with parameters from Table~\ref{tab:tu}. The units are meV.
    }
    \label{fig:3d-epc}
\end{figure}

A strength of this framework is that, being a lattice model, it is not restricted to the small-$k$ regime. We show in Fig.~\ref{fig:3d-epc}(a) the comparison between DFPT and the EPC of the three-band model for a high $E_{F,2}=240$ meV case [indicated also in Fig.~\ref{fig:FS}(a)]. As seen, even when parts of the FS extend close to the BZ edge (along the [100] and symmetry equivalent directions) the agreement between the EPC by the three-band model (dashed lines) and \emph{ab initio} computations is very good.

As a consequence, another notable strength of the EPC model Eq.~\eqref{eq:Hu} is that one can straightforwardly map out the three-dimensional (3D) $\bm k$-space EPC, bypassing the computational effort required by relativistic DFPT and similar \emph{ab initio} methods. 
Figs.~\ref{fig:3d-epc}(b)-(c) show an example of the EPC projected on the 3D FSs for $E_{F,1}=40$ meV and $E_{F,2}=240$ meV, respectively, for a $\bar S_1$ mode polarized along [001] (black arrow). 
The corresponding 2D cuts on the $k_z=0$ plane were shown in Fig.~\ref{fig:qe-vs-tb}(a) and Fig.~\ref{fig:3d-epc}(a). As seen, the FS acquires anisotropy in momentum space as $E_F$ is increased, from the quasi-sphere shape in (b) to the three interpenetrating ellipsoids oriented along [100], [010], and [001] in (c). The structure of the EPC evolves accordingly, showing very large anisotropy in (c), with stronger (weaker) coupling in areas close to (far from) the zone center, where the effect of SOC (kinetic term) dominates \cite{gastiasoro2023}. These types of 3D maps can be used to study phenomena where the EPC may play an important role, such as transport, superconductivity, or ferroelectric states.   

\section{Summary and Conclusions}
\label{sec:conclusions}

We have studied the EPC to the zone-center phonons in KTO, which are odd-parity modes.
We focused mainly on the polar $T_{1u}$ modes, particularly on the soft TO mode,  employing a combination of relativistic DFPT and a three-band microscopic model.
Our computations go beyond the \emph{frozen phonon} method, where only intraband terms can be studied, and we find that interband matrix elements are of the same order of magnitude as intraband elements.
Therefore,  interband terms cannot be \emph{a priori} neglected when addressing KTO, and its importance in a physical process must be assessed on a case-by-case basis.    

The DFPT results can be understood in terms of a three-band model for the spin-orbit coupled $t_{2g}$ orbitals. 
After projecting onto the relevant modes, we can study intraband and interband EPC and disentangle spin-conserving type I and spin-non-conserving type II processes.
The spin-orbit-assisted EPC results to be an isotropic dynamical Rashba effect for band 3 ($j=\frac{1}{2}$) and an anisotropic one for bands 1 and 2 ($j=\frac{3}{2}$). 

We find that, for the soft TO mode ($\bar S_1$), spin-non-conserving hopping terms are as large as spin-conserving terms ($\tau_{u,A}=1.08\, t_u$, $\tau_{u,C}=0.44\, t_u$), and need to be included to obtain the correct form of the anisotropy of the EPC in this system.
Therefore, SOC is important not only in the mixing of orbitals for the electronic band structure, but also in determining their coupling to the phonon modes.
The match between the three-band model and DFPT results is very good for all three modes, particularly for interband matrix elements. The discrepancy found for intraband matrix elements is largest in the EPC of modes with significant spin-non-conserving processes $\bar S_1$ and $\bar S_3$ [Fig.\ref{fig:qe-vs-tb}(a) and (c)]. Since the origin of these terms is relativistic, it could be due to restricting the model to the $t_{2g}$ manifold, and an obvious extension would be to include the $e_g$ manifold. Another possibility would involve sub-leading terms beyond neighboring Ta sites in Eqs.\eqref{eq:tu}-\eqref{eq:tau3}, which, according to the Wannier projection, are individually smaller, but their cumulative role could be quantitatively appreciable.

The three-band model works surprisingly well also beyond small ${\bm k}$. 
As such, it also allows for a 3D map of the $\bm k$-space EPC, which can be useful for the study of phenomena involving EPC, such as transport, superconductivity, or ferroelectricity. Indeed, we find the coupling to the soft TO mode to be rather large, making the Rashba-like coupling a promising pairing mechanism in KTO. The role of interband coupling in the Cooper channel should be considered in this case.

The electron-phonon coupling projected on FS cuts perpendicular to different crystalographic orientations shows anisotropies. However, we did not find an evident dominant feature that could explain the different behavior of exposed sample surfaces regarding superconductivity. That said, one should take into account that a small change in the EPC can be strongly amplified due to the exponential dependence of $T_c$ on the EPC at weak coupling.  Therefore, the explanation of the observed anisotropy on $T_c$ requires detailed computations on the surface that go beyond our present scope. One step in this direction would be to consider surface phonons \cite{chu2024,zhou2025}. 

Our DFPT protocol was built to be general and allow for the extraction of model coefficients for a given phonon mode. By computing, in fact, the first-order coefficients of the Kohn-Sham Hamiltonian for each atom and direction, we can project them onto the appropriate mode basis.
This circumvents the eventual numerical problems arising when computing degenerate phonon modes, which will be, in general, a mixture of the $\bar S_\lambda$ modes. 
This method is quite general and may be applied to any other system with even or odd parity vibrational modes.

\section*{Acknowledgements}
We thank I. Paul, D. Pelc and M. R. Norman for fruitful discussions. M.N.G is supported by the Ramon y Cajal Grant RYC2021-031639-I funded by MCIN/AEI/ 10.13039/501100011033.
G.V. acknowledges support from the Swiss National Science Foundation (SNSF) via Swiss Postdoctoral Fellowship grant number: TMPFP2 224637.
We acknowledge the CINECA award under the ISCRA initiative Grants No. HP10CPHFAR and HP10CMZFMM, for the availability of high-performance computing resources and support.

\appendix

\section{Details of the three-band model}
\label{app:soc-epc}
\subsection{Parameters for electronic bands}
\label{app:appH0}
We include hopping terms up to next-nearest neighboring $t_{2g}$ orbitals in Ta atoms for the tight binding model in Eq.~\eqref{eq:H0},
\begin{align}
\label{eq:tmumu0}
     t^{(0)}_{\mu\mu}(\bm{k})=&-2t_1\left(\cos k_\alpha+ \cos k_\beta \right)-2 t_2 \cos k_\mu \\\nonumber
     &-4 t_3 \cos k_\alpha \cos k_\beta+(4t_1+2t_2+4t_3)\\
   t^{(0)}_{\mu\nu}(\bm{k})=&-4 t_4 \sin k_\mu \sin k_\nu,\label{eq:tmunu0}
 \end{align}
with parameters $t_1=483.9$ meV, $t_2=30.36$ meV, $t_3=196.9$ meV and $t_4=27.6$ meV obtained by fitting the QE electronic structure [Fig.~\ref{fig:FS}(a)]. In Eq.~\eqref{eq:tmumu0} $\alpha\neq\beta\neq\mu$, while in Eq.~\eqref{eq:tmunu0},  $\mu \neq \nu$. The SOC gap at the zone center in Fig.~\ref{fig:FS}(a) is given by $3\xi=415.9$ meV. 
The parameters are overall consistent with those given in Ref. \cite{venditti2023}, whereas the small differences can be ascribed to the slightly larger lattice constant adopted here as well as to different implementations of DFT in QE (used in this paper) and VASP (used in Ref. \cite{venditti2023}).

\subsection{Wannierization}
\label{app:wannier}
We projected the fully relativistic band structure of KTO on maximally localized Wannier functions (MLWFs) \cite{wannier90_RMP} as implemented in Wannier90 \cite{wannier90}. Conduction bands of cubic KTO calculated with QE have been projected on spinor Wannier functions localized on Ta atoms with real angular functions $yz$, $xz$, $xy$, corresponding to spin-1/2 $t_{2g}$ orbitals $\ket{\mu\sigma}$ introduced before, with $\mu=x,y,z$, respectively. 
FS cuts have been then calculated using the Wannier interpolation procedure implemented in Wannier90 \cite{wannier90_RMP,wannier90} along three different slices in the Brillouin zone, perpendicular to [001], [110] and [111] directions; each slide has been sampled on a uniform 100$\times$100 $k$-point grid for two values of the Fermi energy ($E_{F,1}=40$ meV and $E_{F,2}=240$ meV) defined from the conduction band minimum. 

We also construct a real-space tight-binding (TB) Hamiltonian in the basis of MLWFs (MLWF-TB), which automatically includes also terms beyond next-nearest neighbors. Since indirect processes via different orbitals/atoms are effectively included when downfolded to the $t_{2g}$ manifold, SOC effects are accounted for by both a local term (effective atomic SOC) and by complex hopping interactions. SOC-induced effective parameters can be identified by comparing the MLWF-TB Hamiltonians constructed from \emph{ab initio} calculations with and without SOC. The local SOC has the form given in Eq.~\eqref{eq:H0} with a coupling constant $\xi_\mathrm{MLWF}=139.2$ meV, while SOC-induced hopping terms are of the order of a few meV and can be thought of as being effectively absorbed in the effective SOC of the three-band tight binding model Eq. \eqref{eq:H0}. For completeness, we also provide the TB-MLWF parameters up to next-nearest neighbors, $t_1^\mathrm{MLWF}=473.1$ meV, $t_2^\mathrm{MLWF}=23.2$ meV, $t_3^\mathrm{MLWF}=81.5$ meV, and $t_4^\mathrm{MLWF}=13.8$ meV. It is worth mentioning that the TB-MLWF Hamiltonian includes hopping terms beyond next-nearest neighbors, which can be thought to be effectively absorbed in the three-band TB model of Eq.~\eqref{eq:H0}, explaining the different values of TB parameters.

The identification of induced hopping terms arising from polar phonons has been achieved within a $\bm q =\bm 0$ frozen-phonon approach. We constructed three distorted structures following the $\bar{S}_\lambda$ modes along the [001] direction with fixed amplitude $u=0.002$ \AA~ to guarantee linearity of the coupling to polar modes~\cite{venditti2023}. The band structure with and without SOC of each distorted cell has been projected on Ta $t_{2g}$ MLWFs. By analyzing the computed MLWF-TB Hamiltonians, we can single out the odd-parity hopping interactions induced by the lowering of symmetry, as well as their non-relativistic or relativistic origins. Such frozen-phonon induced hopping parameters $\bar{t}_{\mu\nu,\sigma\sigma'}$ can be related to the EPC entering in Eq.~\eqref{eq:Hu} exploiting the linearity of the coupling regime, hence by multiplying $\bar{t}_{\mu\nu,\sigma\sigma'}$ with the ratio $l_{\bm q=\bm 0,\lambda}/u$.

When SOC is turned off, the largest induced hopping for the Slater $\bar{S}_1$ mode is the nearest-neighbor one giving rise to type I EPC parametrized by $t_u$. We find that next-nearest-neighbor processes also contribute to type I EPC: such contributions are subdominant for the $\bar{S}_1$ mode, while for $\bar{S}_2$ and $\bar{S}_3$ modes they are comparable in strength with the NN process. The type I EPC parameters evaluated from non-relativistic MLWFs but taking into account only the nearest-neighbor contribution are $t_u(\bar{S}_1) = 35.9$ meV, $t_u(\bar{S}_2) = 1.7$ meV and $t_u(\bar{S}_3) = 2.6$ meV. Remarkably, they display the same order of magnitude of type I EPC effective parameters $t_u$ evaluated by fitting non-relativistc DFPT results with the model Eq. \eqref{eq:Hukxnosoc} (thus taking into account beyond-NN contributions) and listed in Table \ref{tab:tu}. By including contributions arising from next-nearest-neighbor induced hoppings in the TB-MLWF Hamiltonian,  the type-I EPC parameters are modified as $t_u(\bar{S}_1) = 48.9$ meV, $t_u(\bar{S}_2) = 2.8$ meV and $t_u(\bar{S}_3) = 3.9$ meV. 
Finally, the inclusion of SOC introduces subleading modifications to the type I spin-conserving term, namely $t_u(\bar{S}_1) = 36.1$ meV, $t_u(\bar{S}_2) = 1.7$ meV, and $t_u(\bar{S}_3) = 2.8$ meV, when considering NN hoppings. At the same time, several new odd-parity spin-dependent hopping terms mediated by SOC appear in the distorted structures. 

In order to keep the TB modelization as simple as possible, while still capturing the essential physics qualitatively, we singled out the leading spin-dependent hopping processes between nearest neighbors that are sketched in Fig.~\ref{fig:tu}. They can be interpreted as arising from indirect processes involving spin-orbit-coupled different orbitals/atoms that are downfolded into the spin-1/2 $t_{2g}$-only manifold. For instance, the process sketched in Fig.~\ref{fig:tu}(b) between two Ta atoms across a bridging O atom in the Ta$^L$-O-Ta$^R$ bond may arise from SOC mixing O-$p$ states with opposite spin via the sequence $d_{yz,\uparrow}^L\xrightarrow[\mbox{\scriptsize i-hop}]{} p_{y,\uparrow} \xrightarrow[L_xS_x]{} p_{z,\downarrow} \xrightarrow[\mbox{\scriptsize hop}]{} d_{yz,\downarrow}^R$, where we label i-hop and hop the hopping processes between Ta-$d$ and O-$p$ states induced or not by the polar distortion. An alternative indirect process contributing to the same effective matrix element but involving Ta-$e_g$ states can occur along the sequence $d_{yz,\uparrow}^L\xrightarrow[\mbox{\scriptsize i-hop}]{} p_{y,\uparrow} \xrightarrow[\mbox{\scriptsize hop}]{} d^R_{x^2-y^2,\uparrow} \xrightarrow[L_xS_x]{}d^R_{yz,\downarrow}$. The leading effective spin-dependent hopping processes induced by the three polar modes are parametrized by three independent coefficients $\tau_{u,A},\tau_{u,B},\tau_{u,C}$ for each mode, listed in Table \ref{tab:tu} and given in units of the spin-independent induced hopping $t_u$.

\subsection{Extract type I parameter from non-relativistic DFPT}
\label{app:tu}

We can do the same analysis as in Subsection~\ref{sec:nosoc} but for $\bm k$ along [110]. In this case, the three-band hopping Hamiltonian reads
\begin{flalign}
    &h_0(\xi=0,\bm k_M)=\sum_{i=1,2,3} \varphi^\dagger_i(\bm k_M) E_i(\bm k_M)\sigma_0\varphi_i(\bm k_M)&
\end{flalign}
with energy dispersions
\begin{flalign}
    &E_1(\bm k_M)=4\left[t_1+2t_3+t_2-2t_4(1-\cos \frac{k}{\sqrt{2}})\right] \sin^2\frac{k}{2\sqrt{2}}&\nonumber\\
    &E_2(\bm k_M)=4\left[t_1+2t_3+t_2+2t_4(1-\cos k)\right] \sin^2\frac{k}{2\sqrt{2}}&\nonumber\\
    &E_3(\bm k_M)=8\left[t_1+t_3(1+\cos \frac{k}{\sqrt{2}})\right]\sin^2\frac{k}{2\sqrt{2}}&
\end{flalign}
Along this $\bm k$ direction, we have three doubly spin-degenerate bands. Considering the same spin-independent inter-orbital matrix elements induced in the presence of a polar phonon, we obtain again pure \emph{interband} terms for the type I model:
\begin{flalign}
    \label{eq:Hukmnosoc}
   & h_\mathrm{EPC}(\bm{k}_M)=\left[-2it_{u}\sqrt{2}\sin(\frac{k}{\sqrt{2}})\varphi_1^\dagger(\bm{k}_M)\sigma_0\varphi_3(\bm{k}_M)\right] \mathcal{A}_{q=0}&\nonumber\\
   &\quad+\mathrm{h.c.} =\varphi_1^\dagger(\bm{k}_M)\Lambda_{13}(\bm{k}_M)\varphi_3(\bm{k}_M)\mathcal{A}_{q=0}+\mathrm{h.c.}&
\end{flalign}
which is analogous to the result in Eq.~\eqref{eq:Hukxnosoc}, i.e. interband hopping terms that preserve spin degeneracy.

\begin{figure}
    \centering
    \includegraphics[width=0.5\textwidth]{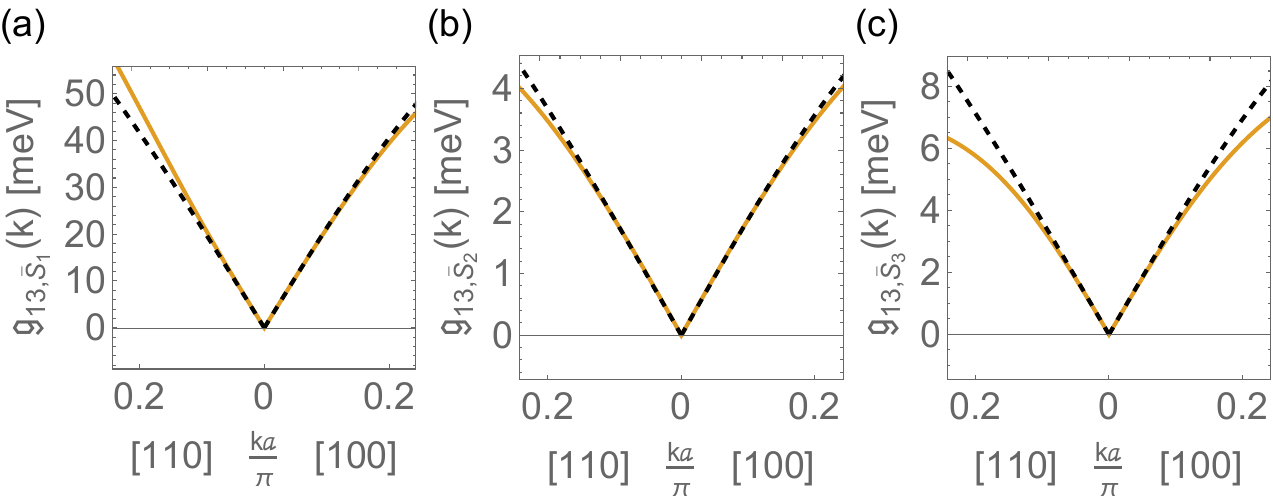}
    \caption{EPC interband term from non-relativistic DFPT computation along high-symmetry-direction (orange lines) for polar modes (a) $\bar S_1$ (b) $\bar S_2$, and (c) $\bar S_3$ polarized along [001].  Fitting the small $k$ region to Eqs.~\eqref{eq:Hukxnosoc} and Eqs.~\eqref{eq:Hukmnosoc} (black dashed lines) we obtain the value for $t_{u}$, listed in Table~\ref{tab:tu}.
    }
    \label{fig:Hu-MGX}
\end{figure}

Doing a non-relativistic DFPT computation in KTO for a polar $\bar S_i$ mode ($i=1,2,3$) polarized along 001, we indeed obtain interband terms along $\bm{k}=\bm{k}_X$ and $\bm{k}=\bm{k}_M$ within DFPT, shown in Fig.~\ref{fig:Hu-MGX} (orange line) for the three polar $\bar{S}_i$ modes. From the fit of these results at small $k$ with Eqs.~\eqref{eq:Hukxnosoc} and Eqs.~\eqref{eq:Hukmnosoc} we extract the type I induced amplitude $t_u$ for each mode, listed in Table ~\ref{tab:tu}.

\section{Computational details}
\label{app:compdet}

\subsection{Density functional theory calculations}

\begin{figure}
    \centering
    \includegraphics[width=0.9\linewidth]{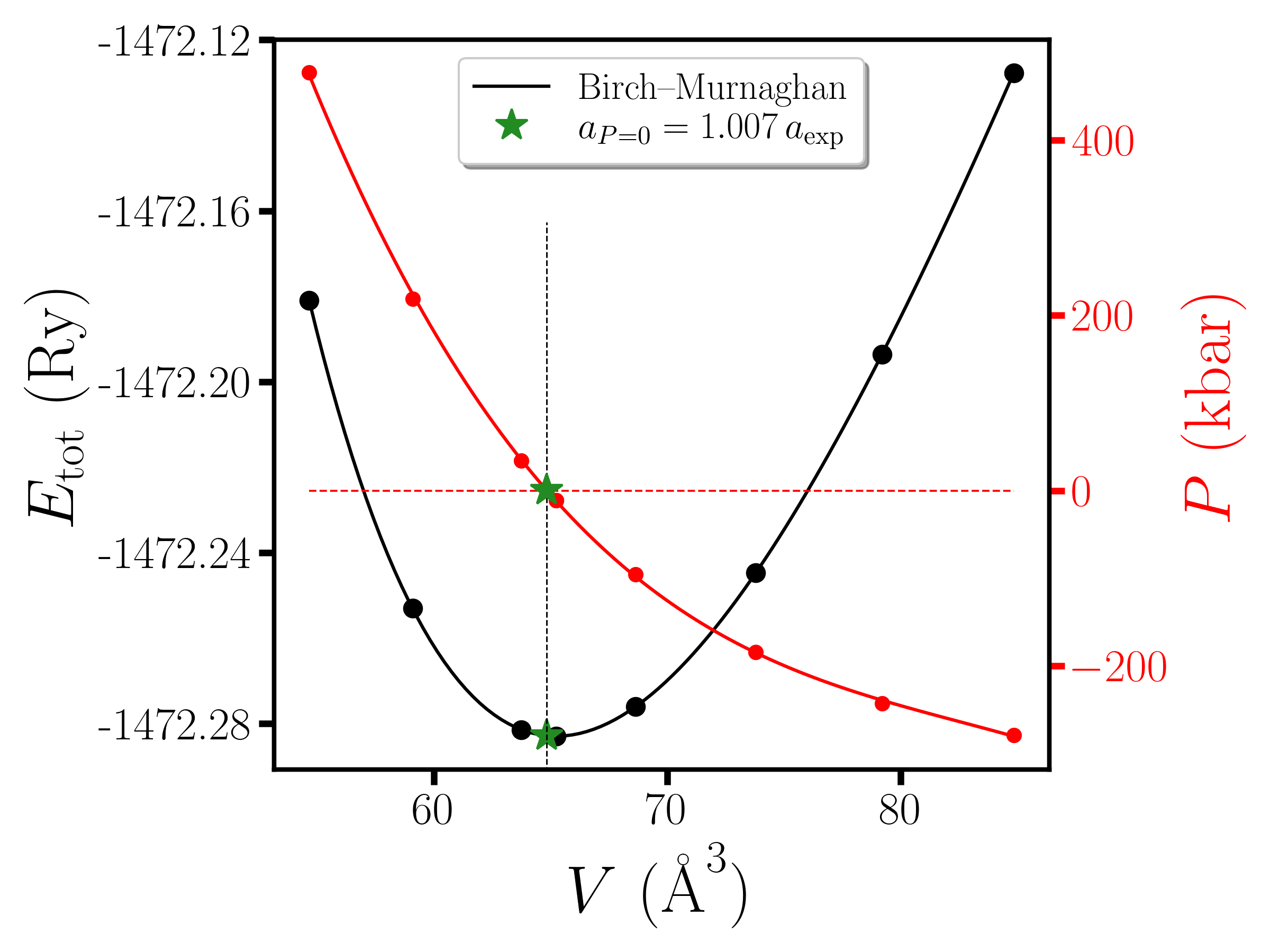}
    \caption{{\em Relaxation cell procedure.} Total energy $E_\text{tot}$ computed within the QE scf calculation as a function of volume $V=a^3$ (left axis) and corresponding pressure $P$ (right axis, red). 
    Solid lines are the Birch-Murnaghan energy potential fit of $E_\mathrm{tot}$ (black) and its derivative ($P$, red).
    We take the value $a_{P=0}=1.007\,a_\text{exp}$ (green stars), corresponding to an ideal zero pressure. }
    \label{fig:en_vs_vol}
\end{figure}

We first computed the optimum value of the cell parameter $a$ of KTO.
In Fig.~\ref{fig:en_vs_vol} we show the computed total energy $E_\mathrm{tot}$ of KTO as a function of volume $V=a^3$ (left axis, black) and the corresponding computed pressures $P$ (right axis, red).
We fitted the curve $E_\mathrm{tot} $with the Birch-Murnaghan potential \cite{Birch1947} and $P$ with its derivative.
We then took the cell parameter corresponding to zero pressure, indicated by $a_{P=0}$ in the plot.
We found $a_{P=0}=1.007\,a_\mathrm{exp}=4.0175\,\text{\AA}$ (green stars), where $a_\text{exp}=3.9885\,$\AA\, is the experimental value~\cite{Vousden1951}.
This relaxation procedure was done using an automatic mesh $10\times 10 \times 10$, checking that the results for and $a_{P=0}$ were unchanged using a tighter grid of $14\times 14 \times 14$.
In both cases, $E_\text{tot}$ remained the same up to the 6$^\text{th}$ digit (in Ry units) and the pressure computed in QE at $a_{P=0}$ was found to be $P=0.56\,$kbar, low enough to have a reliable convergence in optical phonons.   
In all our calculations we then used $a_{P=0}$ 
as the cubic cell parameter and an automatic mesh  $10\times 10 \times 10$ for the self-consistent calculations.

To further ease the calculations, we give as input in the self-consistent QE calculation the 56 $k$-points with the correct weights corresponding to the automatic $10\times 10 \times 10$ automatic grid, plus the $k-$path (with zero weight) along which we want to compute the EPC:
 high-symmetry-lines $M-\Gamma-X$ are given with a total of 400 equidistant points, FS are instead computed via the Wannier interpolation procedure 
 explained in Appendix \ref{app:wannier}.

We used fully relativistic (scalar relativistic) PAW pseudopotentials for calculations with SOC (without SOC) \cite{DALCORSO2014337} and the generalized gradient approximation for the exchange and correlation functional (PBE) \cite{PBE}.
Convergence thresholds on total energy and forces were taken respectively as $5\cdot 10^{-6}$ and $10^{-5}$.
  Cutoffs on wavefunctions and charge densities were respectively 55 and 550 Ry. 
Convergence threshold for self-consistency of electrons was $10^{-12}$, with a maximum number of iterations for step 80 and mixing factor 0.4.
The threshold for self-consistency in the phonon computation was $10^{-14}$.

For electron-phonon calculations, we used an in-house-modified version of QE, which allows for the evaluation of Eq.~\eqref{eq:avrg} for any choice of the phonon eigenvector. In particular, we print the electron-phonon on the cartesian basis set, i.e., the $\gamma^{\kappa\alpha}_{nm}(\bm k, \bm q)$ of Eq.~\eqref{eq:LambdaQE}. This is obtained by linking the displacements of each irreducible representation to the Cartesian one. These tasks are all performed by the \texttt{ph.x} program, which first performs the standard DFPT calculation of the linear response and, after convergence, prints the electron-phonon. The contraction of the $\gamma^{\kappa\alpha}_{nm}(\bm k, \bm q)$ matrices with any given phonon eigenvector is then easily performed in postprocessing.

\subsection{Eigenfrequencies and eigenmodes}
\label{app:QEdecomposition}

Decomposing the eigenvectors $\bm e^\lambda_\kappa(q=0)$ computed by QE in the basis ($\bar S_1$, $\bar S_2$, $\bar S_3$, $\bar S_4$) we get the following coefficients
 \begin{flalign*}
& T_{1u}:\,\, (0.966, 0.034, 0, 0) &(4.2\,\mathrm{meV})  \\
& T_{1u}:\,\, (0.034, 0.965, 0, 0) & (22.7 \,\mathrm{meV}) \\
& T_{1u}:\,\, (0, 0, 1, 0) &(63.5 \,\mathrm{meV})\\
& T_{2u}:\,\, (0, 0, 0, 1) &( 32.3\,\mathrm{meV})
 \end{flalign*}
where we have also added their corresponding frequency in meV.

The eigenvectors and frequencies of the TO phonons determined by hyper-Raman scattering experiments at room temperature~\cite{Vogt1988}:
\begin{flalign*}
    &T_{1u}:\,\,{(0.991, 0.012, 0.0027, 0.)}&(10.0\,\,\mathrm{meV})\\
    &T_{1u}:\,\,{(0.011, 0.994, 0.,0.)}&(24.7\,\,\mathrm{meV})\\
    &T_{1u}:\,\,{(0.002, 0, 0.998, 0.
)}&(67.7\,\,\mathrm{meV})\\
    &T_{2u}:\,\,{(0, 0, 0, 1)}&(34.6\,\,\mathrm{meV})
\end{flalign*}
Therefore, the phonon spectra at the zone center computed by DFPT 
give results very similar to the experimentally reported eigenvectors (nearly pure $\bar S_\lambda$ modes) and frequencies.
When computing Eq.~\eqref{eq:LambdaQE} we used the experimental frequencies in Ref.~\cite{Vogt1988} except for the soft TO mode, which has a strong temperature dependence and we used instead its low temperature experimental frequency 2.5 meV, reported in Refs.~\cite{vogt1995,chu2024}.

\begin{figure*}[th]
    \centering
\includegraphics[width=\textwidth]{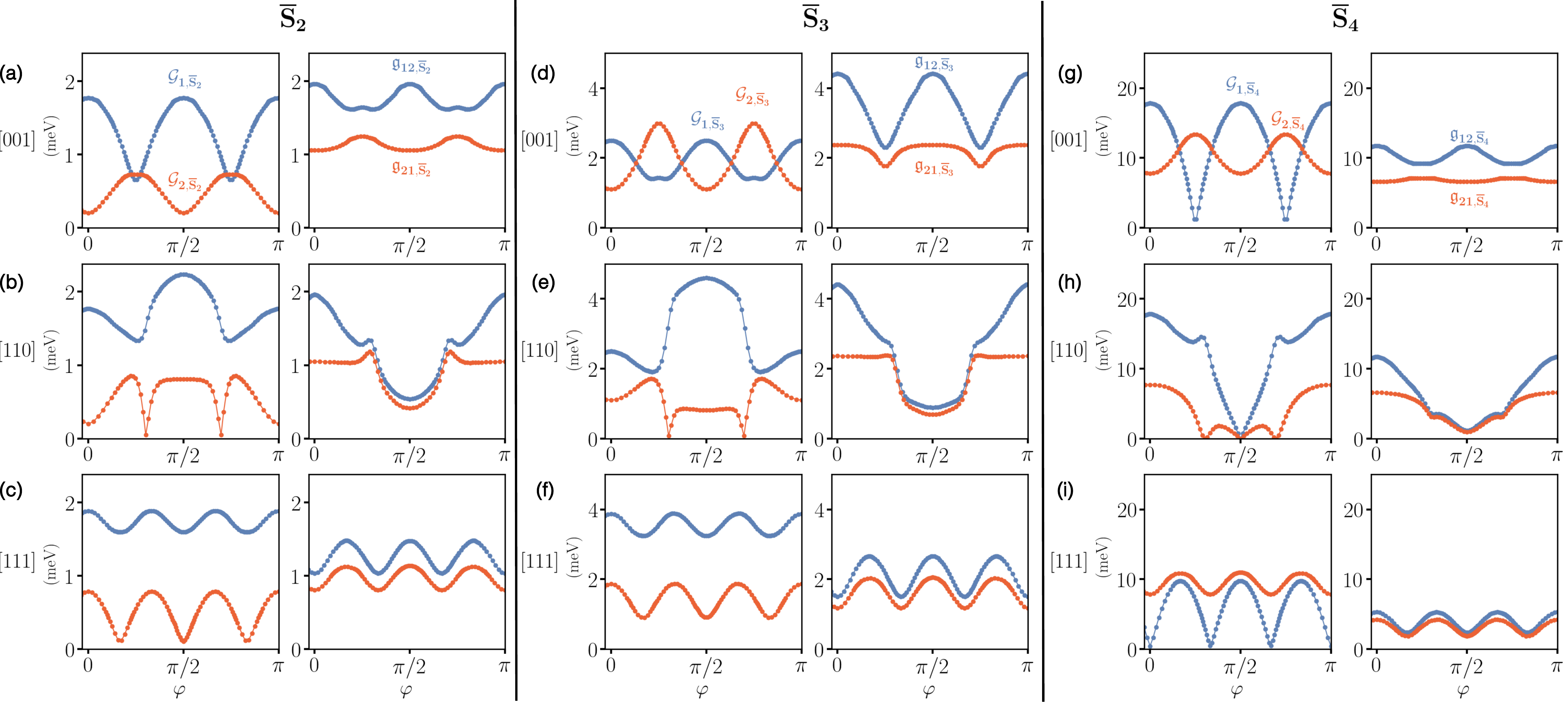}
    \caption{Orientation dependence of the $\bm q=\bm 0$ intra- and inter-band EPC for the following modes polarized perpendicular to the surface examined: (a)-(c) $\bar S_2$, (d)-(f) $\bar S_3$, and (g)-(i) $\bar S_4$. Panels are labeled on the left by the polarization of the mode. The EPC is evaluated on the Fermi surface with $E_{F,1}=40$ meV. $\varphi$ is the azimuthal angle along the Fermi surface cut. 
    We used the experimental values for the frequency $\omega_{\bm q=0, \lambda}=24.7, 67.7,\, 34.6$\,meV respectively~\cite{vogt1995}. }
    \label{fig:epcQE-S2-S3-S4}
\end{figure*}

\subsection{Check against frozen phonon calculations}
\label{app:frph}
The intraband DFPT results match the EPC obtained by a frozen phonon method~\cite{gastiasoro2023,venditti2023}. For that, we compute the relativistic electronic band splitting 
$\delta E_n (\bm k)$ in the presence of a polar distortion $\bar S_\lambda$ with amplitude $u$, which gives the following EPC 
\begin{equation}
\mc{G}^{\mathrm{fr.ph.}}_{n,\lambda}(\bm k)=\frac{\delta E_{n,\lambda}(\bm k) }{2u}\sqrt{\frac{\hbar}{2 \mu_{\lambda}\omega_{0}}} 
   \label{eq:gfrozen}
\end{equation} 
The result is in agreement with degenerate perturbation theory to first order in $u$, i.e., interband processes can only affect higher order in $u$ quantities, not the breaking of the degeneracy. Hence, to linear-in-$u$, the EPC obtained from DFPT and frozen phonon are formally exact, and we find this to be respected up to a precision of 0.005$\div$0.05 meV (root mean square of $\lesssim 0.01$ meV) for displacements of $u=0.001\,$\AA.

\subsection{Orientation dependence for optical modes}
\label{app:DFPT_S2_S3}

We show for completeness our DFPT results for modes $\bar S_2,\,\bar S_3,\,\bar S_4$ in Fig.~\ref{fig:epcQE-S2-S3-S4} for FS cuts at $E_{F,1}=40$\,meV perpendicular to the three orientations [001], [110], and [111].
The strength of the EPC has a strong azimuthal angular dependence in all cases, as seen for the $\bar S_1$ mode [Fig.~\ref{fig:epc-phi}]. 
Similar to the soft mode, the strength observed along the three FS cuts are comparable in all modes, making it difficult to claim anything from our calculations about the orientation dependence of $T_c$.
As we underline in our conclusions, however, in the weak coupling regime small differences are strongly enhanced by the exponential dependence.
A detailed computation of $T_c$ goes beyond the scope of this work and is left for future investigations.

\bibliography{biblio}

\end{document}